\documentclass[a4paper,fleqn,usenatbib]{mnras}

\usepackage{newtxtext,newtxmath}

\usepackage[T1]{fontenc}
\usepackage{ae,aecompl}
\usepackage{natbib}
\usepackage{xcolor}

\usepackage{graphicx}	
\usepackage{amsmath}	
\usepackage[utf8]{inputenc}
\DeclareUnicodeCharacter{2212}{-}
\usepackage[english]{babel}
\usepackage{tabu}
\usepackage{subfloat}
\usepackage{breakurl}
\usepackage[justification=centering]{caption}
\usepackage[]{threeparttable}
\usepackage{float}
\usepackage{stfloats}

\title[{\it Spectral and temporal features of LMC X-1, LMC X-3}]{Broadband `spectro-temporal' features of extragalactic black hole binaries LMC X-1 and LMC X-3: An {\it AstroSat} perspective}

\author[Bhuvana et al.]{
Bhuvana G. R.$^{1}$\thanks{E-mail: bhuvanahebbar@gmail.com}, Radhika D.$^{1}$, V. K. Agrawal$^{2}$, Samir Mandal$^{3}$ and Anuj Nandi$^{2}$ 
\\
$^{1}$Department of Physics, Dayananda Sagar University, Hosur Main Road, Bengaluru, 560068, India.\\
$^{2}$Space Astronomy Group, ISITE Campus, U. R. Rao Satellite Center, Outer Ring Road, Marathahalli, Bengaluru, 560037, India.\\
$^{3}$Department of Earth and Space sciences, Indian Institute of Space Science and Technology, Valiamala, Thiruvananthapuram, 695547, India.\\
}

\date{Accepted XXX. Received YYY; in original form ZZZ}

\pubyear{2019}

\begin{document}
\label{firstpage}
\pagerange{\pageref{firstpage}--\pageref{lastpage}}
\maketitle

\begin{abstract}
	We present the first results of extragalactic black hole X-ray binaries LMC X-1 and LMC X-3 using all the archival and legacy  observations by {\it AstroSat} during the period of $2016-2020$. Broadband energy spectra ($0.5-20$ keV) of both sources obtained from the \textit{SXT} and \textit{LAXPC} on-board \textit{AstroSat} are characterized by strong thermal disc blackbody component ($kT_{in}\sim1$keV, $f_{disc}>79\%$) along with a steep power-law ($\Gamma\sim2.4-3.2$). Bolometric luminosity of LMC X-1 varies from $7-10\%$ of Eddington luminosity ($L_{Edd}$) and for LMC X-3 is in the range $7-13\%$ of $L_{Edd}$. We study the long-term variation of light curve using \textit{MAXI} data and find the fractional variance to be $\sim25\%$ for LMC X-1 and $\sim53\%$ for LMC X-3. We examine the temporal properties of both sources and obtain fractional rms variability of PDS in the frequency range $0.002-10$ Hz to be $\sim9\%-17\%$ for LMC X-1, and $\sim7\%-11\%$ for LMC X-3. The `spectro-temporal' properties indicate both sources are in thermally dominated soft state. By modelling the spectra with  relativistic accretion disc model, we determine the mass of LMC X-1 and LMC X-3 in the range $7.64-10.00$ $M_{\odot}$ and $5.35-6.22$ $M_{\odot}$ respectively. We also constrain the spin of LMC X-1 to be in the range $0.82-0.92$ and that of LMC X-3 in $0.22-0.41$ with 90\% confidence. We discuss the implications of our results in the context of accretion dynamics around the black hole binaries and compare it with the previous findings of both sources. 
\end{abstract}

\begin{keywords}
X-ray binaries -- accretion, accretion discs-- black hole physics -- stars: black holes -- radiation mechanisms: general -- stars: individual: LMC X-1 -- stars: individual: LMC X-3
\end{keywords}



\section{Introduction}
\label{Intro}
Black Hole X-ray Binaries (BH-XRBs) consist of a black hole (BH) along with a normal companion star which are gravitationally bound to each other. Highly compact BH accretes matter from the companion star and forms an accretion disc around it. The accretion disc gets heated up due to the geometrical compression in the accretion process and gives out energy in different wavebands, mainly in X-rays. Based on the type of X-ray emission, XRBs are classified either as persistent sources or as transients \citep{1997ApJ...491..312C,2016ApJS..222...15T,2016A&A...587A..61C}. Persistent sources are the ones that radiate X-ray luminosity consistently for a long period of time \citep{1996ARA&A..34..607T}. Transients, on the other hand, undergo outburst emitting large X-ray flux \citep{1989ApJ...337L..81T} once in a while. Generally, XRBs exist either in a low mass X-ray binary (LMXB) system or high mass X-ray binary (HMXB) system \citep{2006ARA&A..44...49R} depending on the mass of the companion star. Most of the observed transient sources typically exist in LMXBs whereas persistent sources are found in HMXBs.\par
To study the accretion dynamics around the black holes, it is necessary to understand the X-ray spectrum of the source.
The radiation spectrum of BH-XRBs generally consists of thermal and non-thermal components. While thermal radiation comes from different radii of accretion disc \citep{1973A&A....24..337S}, non-thermal emission is due to the Comptonization of photons from the disc by hot corona \citep{1995xrbi.nasa..126T,1995ApJ...455..623C}. Thermal radiation from the disc forms the low energy soft X-ray flux in the energy spectra and non-thermal radiation contributes to the high energy. The emitted radiation varies with respect to energy and time, resulting in different spectral states. These spectral states are usually classified as low/hard state (LHS), hard-intermediate state (HIMS), soft-intermediate state (SIMS) and high/soft state (HSS) (\citealt{2001ApJS..132..377H,2005AIPC..797..197B,2006ARA&A..44...49R, 2012A&A...542A..56N,2014AdSpR..54.1678R,2016MNRAS.462.1834R,2018Ap&SS.363...90N,2019MNRAS.487..928S,2020MNRAS.tmp.2075B, Katoch2020} (under review) and references therein). Persistent BH-XRBs mainly exhibit HSS and LHS \citep{1973ApJ...180..531S, 1986ApJ...308..110M,1999ApJ...525..901M,2002ApJ...578..357Z,2006ARA&A..44...49R}. During HSS, the X-ray energy spectrum of the source consists of a strong disc component and a weak power-law component. While the source is in LHS, the spectrum consists of a hard power-law component \citep{1995ApJ...455..623C,1995xrbi.nasa..126T,1996ApJ...466..404C,2001ApJS..133..187H,2002ApJ...578..357Z,2012A&A...542A..56N,2018JApA...39....5S,2019MNRAS.486.2705A}. \par
In addition to the spectral variability exhibited at different states, BH-XRBs show timing variability. This variability can be understood by studying the source light curve and Power Density Spectrum (PDS) which allows us to investigate the variation in power at different frequencies. During LHS, the PDS can be described by a flat-top noise with a broken power-law, and sometimes have signatures of Quasi-periodic Oscillations (QPOs) \citep{2005ApJ...629..403C,2006ARA&A..44...49R,2011BASI...39..409B}. PDS during HSS is characterized by a weak power-law component. Narrow frequency features are rarely seen during this state. The fractional rms amplitude in the source light curve is seen to be high ($>10\%$) during LHS \citep{1990A&A...230..103B,1995xrbi.nasa..126T,1995PASP..107.1207N,2012A&A...542A..56N} and it decreases ($<10\%$) as the source moves towards the HSS.
\par
During HSS, the inner edge of the disc moves towards the BH and it is predicted to be very close to the radius of innermost stable circular orbit i.e. ISCO \citep{1983bhwd.book.....S}. Most of the thermal emission during this state comes from the innermost region of the disc where there is a strong gravitational effect and these soft X-rays gives us information on specific angular momentum of the BH. Angular momentum of the BH can be expressed in terms of a dimensionless parameter called spin as $a=cJ/GM_{BH}^{2}$ whose magnitude varies between 0 and 1 \citep{1997ApJ...477L..95Z,2006ApJ...636L.113S,2006ApJ...652..518M, 2014SSRv..183..295M}. Here, $J$ represents the angular momentum, $c$ is the speed of light, $G$ is gravitational constant and $M_{BH}$ is mass of the BH. Thus by studying the disc component during thermal dominated spectra of a BH, $J$ can be calculated which in turn gives the spin parameter if $M_{BH}$ is known. This method of determining the spin of BH by modelling the broadband energy spectrum is known as the continuum-fitting method \citep{1997ApJ...482L.155Z, 2011CQGra..28k4009M, 2014ApJ...793L..29S}. 
\par
LMC X-1 is the first extragalactic X-ray source found in the Large Magellanic Cloud (LMC) \citep{mark1969}. The system is a HMXB containing an O7 III star along with a black hole, which are orbiting with a period of $\sim$3.9 days \citep{1995PASP..107..145C}. The mass of black hole, inclination angle of the system and distance to the source are estimated to be $10.91 \pm 1.41$\(M_\odot\), $36.38 \pm 1.92^{\circ}$ and $48.1\pm{2.22}$ kpc respectively \citep{2009ApJ...697..573O}. Continuum fitting of energy spectra during the thermal dominated states has shown that it is a rapidly spinning black hole (see also \citealt{2020ApJ...897...84T}) with spin parameter $a=0.92^{+0.05}_{-0.07}$ \citep{2009ApJ...701.1076G}. The X-ray energy spectrum of the source is always found to be in thermal dominated state \citep{1989PASJ...41..519E,1993ApJ...403..684E,1999AJ....117.1292S} with a power-law tail component above $10$ keV. X-ray flux of the source is moderately variable in short period (< 1 ks) \citep{2001MNRAS.320..316N} whereas very stable on long time-scale \citep{2009ApJ...697..573O}. 
The temporal properties of LMC X-1 are similar to that of typical HSS with its PDS approximately following a power-law ($\propto\nu^{-1}$). QPOs of frequency $\sim26-29$ mHz have been reported earlier and seems to be peculiar since such type of QPOs are usually found in hard state \citep{2014MNRAS.445.4259A}. 
\par
LMC X-3 is another persistent, bright, X-ray source in LMC consisting of a Roche lobe filling HMXB \citep{2014ApJ...794..154O} with a BH of mass $6.98\pm{0.56} M_{\odot}$ \citep{2001A&A...365L.273S, 2014ApJ...794..154O}. Studies done on this object show that LMC X-3 is spinning at a very low rate with $a=0.25^{+0.20}_{-0.29}$ \citep{2010ApJ...718L.117S} and the disc inclination angle is $69.24\pm0.72^{\circ}$ \citep{2014ApJ...794..154O}. The source is mostly found in thermal dominated state \citep{2001MNRAS.320..316N} with the occasional transition to LHS \citep{2000ApJ...542L.127B, 2001MNRAS.320..327W, 2012ApJ...756..146S}. LMC X-3 is also found to have undergone an Anomalous Low/Hard State (ALS) repeatedly where the X-ray flux drops to a very low value ($\sim 1 \times 10^{35}$ erg/s) in lower energy  \citep{2017ApJ...849...32T}. The long term variability of the source has a high amplitude of the factor of $\sim$ 4 on a $100-200$ day time scale \citep{1991ApJ...381..526C}. However, the variability in the short term scale is very less of only a few percentages \citep{2001MNRAS.320..316N}. Temporal studies done on the source show that the LHS consists of a strong broadband variability with fractional rms of $\sim$40\% along with a QPO at 0.4 Hz \citep{2000ApJ...542L.127B} and HSS has less variability with no QPOs \citep{1988ApJ...325..119T}. Despite having large flux variation in long term scale, the inner disc radius is found to remain constant over a long period of time \citep{2010ApJ...718L.117S}. 
\par  
In this work, we present the first results of the broadband spectral and temporal study of the extragalactic BH-XRBs LMC X-1 and LMC X-3 carried out using all the \textit{AstroSat} \citep{2001ASPC..251..512A} archival and legacy observations. \textit{AstroSat} with its unprecedented spectral and timing resolution along with its broadband coverage forms an excellent observatory to carry out studies on XRBs. Therefore in this study, we make use of the remarkable characteristics of \textit{AstroSat} to perform the spectral and temporal studies of the sources. Source properties during different epochs which are separated by a time gap of a few months spanning over $\sim4.5$ years of the {\it AstroSat} era are explored. We look into the evolution of the source light curve on long term and short term in order to understand the evolution of fractional variance. While long term variability of the source light curve is studied using \textit{MAXI} data, \textit{LAXPC} observations are used to study the short term variability. We also investigate the nature of the PDS to estimate the fractional rms amplitude and look for the presence of any QPOs by means of temporal analysis. Further, we attempt to constrain the BH mass, spin and accretion rate by applying the relativistic disc model to the broadband energy spectra using continuum fitting method. \par
This paper is organized as follows. In section \ref{odr}, we present the observation and data reduction method of \textit{AstroSat} data. In section \ref{AM}, we discuss the methods of timing analysis and spectral modeling. The results of temporal properties, spectral properties along with the estimation of physical parameters are presented in section \ref{res}. Finally in section \ref{DC}, we discuss and conclude the results of our study of LMC X-1 and LMC X-3 using \textit{AstroSat} observations.
\section{OBSERVATION AND DATA REDUCTION}
\label{odr}
\noindent
Persistent X-ray binaries LMC X-1 and LMC X-3 are observed by \textit{AstroSat} during the period of $2016-2020$ with its X-ray instruments \textit{Soft X-ray Telescope (SXT)} \citep{2017JApA...38...29S} and \textit{Large Area X-ray Proportional Counter (LAXPC)}  \citep{2016SPIE.9905E..1DY,2017caantialibration} in the energy range $0.3-8$ keV and $3-80$ keV respectively. We make use of its broadband energy coverage and excellent timing resolution (10 $\mu$s) of \textit{LAXPC} instrument in our study. We obtain all the archival and legacy observation data of these sources from the {\it AstroSat}-Indian Space Science Data Centre (ISSDC)\footnote{\url{https://astrobrowse.issdc.gov.in/astro_archive/archive/Home.jsp}} (see Table \ref{table1}). We also make use of \textit{MAXI} data obtained from ``MAXI/GSC on-demand web interface''\footnote{\url{http://maxi.riken.jp/mxondem/}} to plot the long term light curve and hardness ratio.
\begin{table}
\caption{The log of LMC X-1 and LMC X-3 observations by \textit{AstroSat}.}
\label{table1}
\begin{tabular}{cccc}
\hline
\multicolumn{4}{c}{LMC X-1}                                                                            \\ \hline
Observation Id          & \begin{tabular}[c]{@{}c@{}}Date\\  (MJD)\end{tabular} & Epoch & Exposure (s) \\ \hline
A02\_118T01\_9000000826 & 57717                                            & 1     & 62210        \\
A03\_119T02\_9000001496 & 57993                                            & 2     & 14554        \\ 
A08\_008T01\_9000003414 & 58854											   & 3	   & 62074		\\
A08\_008T01\_9000003596 & 58939											   & 4		& 4573		\\
A08\_008T01\_9000003596	& 58940												& 5		& 53504		\\	   	\hline
\multicolumn{4}{c}{LMC X-3}                                                                            \\ \hline
Observation Id          & \begin{tabular}[c]{@{}c@{}}Date\\ (MJD)\end{tabular}  & Epoch & Exposure (s) \\ \hline
G05\_212T01\_9000000600 & 57614                                                 & 1     &  25712        \\
G05\_212T01\_9000000672 & 57650                                                 & 2     & 21423        \\
A03\_106T01\_9000001130 & 57846                                                 & 3     & 19408        \\
A03\_106T01\_9000001190 & 57862                                                 & 4     & 28600         \\
A04\_112T01\_9000001698 & 58074                                                 & 5     & 28905        \\
A04\_112T01\_9000001778 & 58104                                                 & 6     & 42962        \\
A04\_112T01\_9000001884 & 58155                                                 & 7     & 32550        \\
A04\_112T01\_9000001908 & 58169                                                 & 8     & 32738         \\
A08\_008T02\_9000003430 & 58860													& 9 	& 56862			\\ \hline
\end{tabular}
\end{table}

\subsection{\textit{SXT} Data Reduction}
\label{sdr}

\begin{figure*}
\centering
\includegraphics[width=5cm]{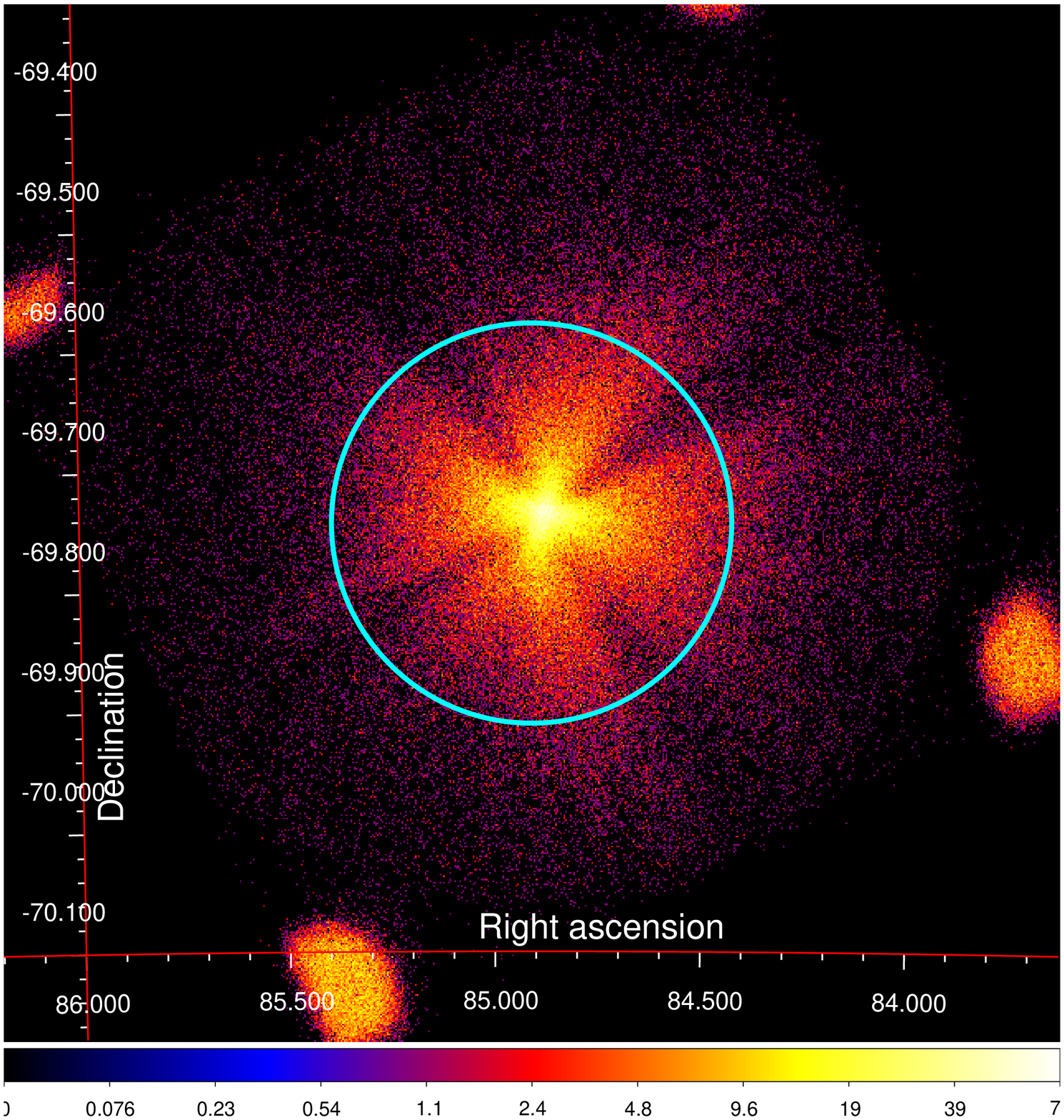} \hspace{1cm}
\includegraphics[width=5cm]{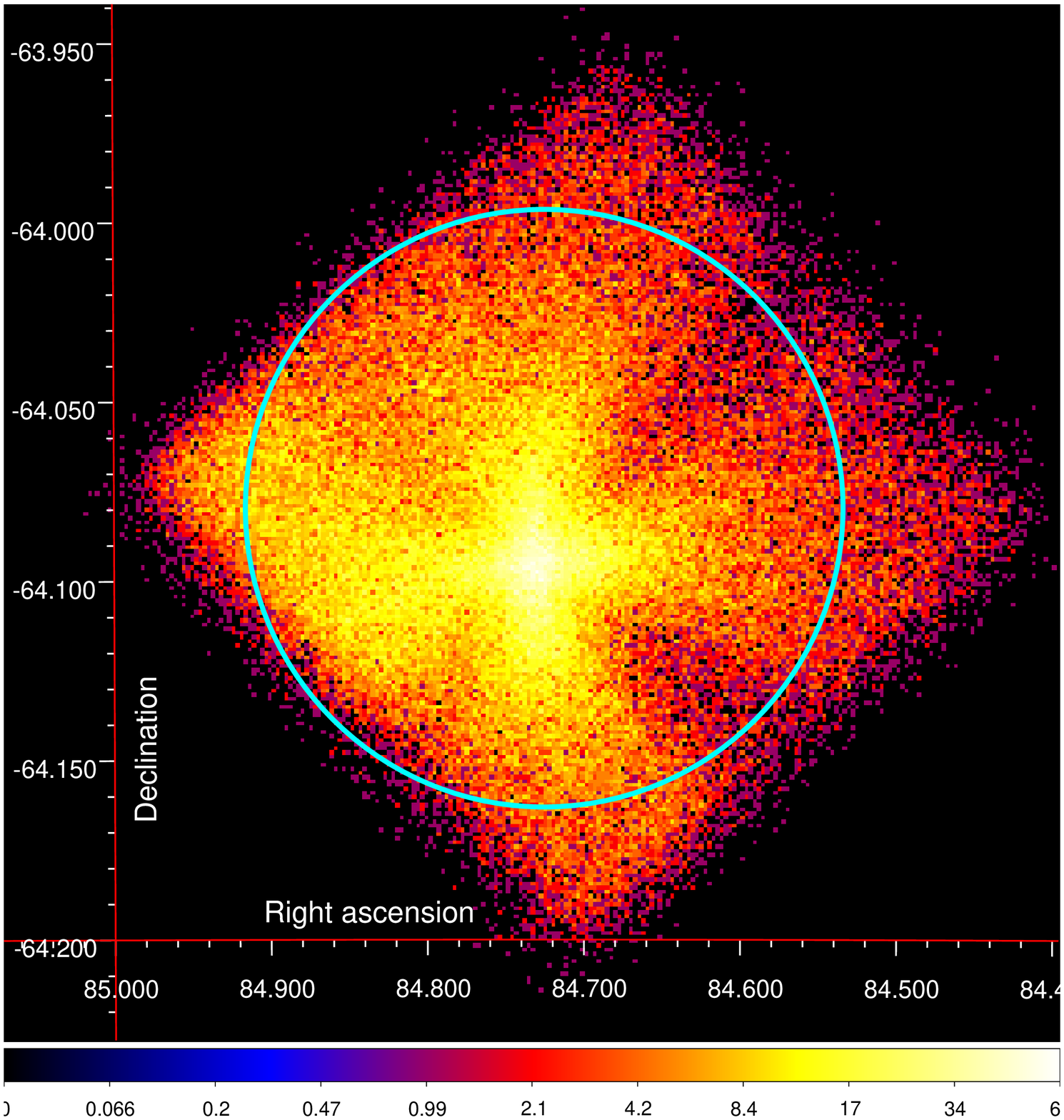}
	\caption{{\it AstroSat - SXT} images of the sources LMC X-1 and LMC X-3 obtained using PC and FW modes are shown respectively in the left and right panels. These two images represent observations during MJD 57717 and MJD 58155 respectively. We consider a circular region of $10\arcmin$ for LMC X-1 and $5\arcmin$ for LMC X-3 as marked in the images from which the spectrum and light curve are extracted (see section \ref{sdr} for details). The bright spots in the corner of PC mode image are from the calibration sources.}
\label{fig1}
\end{figure*}
X-ray imaging instrument \textit{SXT} operates in the energy range of  $0.3-8$ keV.  For the analysis of \textit{SXT} data, we follow the guidelines provided by {\it SXT} team\footnote{\url{https://www.tifr.res.in/~astrosat_sxt/dataanalysis.html}} (see also \citealt{2019MNRAS.487..928S,2020MNRAS.tmp.2075B} for details). \textit{SXT} has observed both sources LMC X-1 and LMC X-3 in Photon Counting (PC) mode and some of the LMC X-3 observations in Fast Window (FW) mode. Level-2 \textit{SXT} data for all the observations are obtained from the ISSDC data archive. Individual orbit data are merged into a single event file with the \textit{SXT} event merger tool using {\it Julia v1.1}. The resultant cleaned event file is used to extract the spectrum and light curve in $0.3-8$ keV energy range using {\it XSELECT v2.5g}. \textit{SXT} image of the source has count rate $< 40$ counts/s and we did not find any effect of pile-up (following the criteria given in {\it AstroSat} handbook\footnote{\url{https://www.issdc.gov.in/docs/as1/AstroSat-Handbook-v1.10.pdf}}). Therefore, a circular region of radius $10\arcmin$ (left panel of Figure \ref{fig1}) and $5\arcmin$ (right panel of Figure \ref{fig1}) are chosen for PC and FW modes respectively. From this region, the spectrum and light curve for further analysis are extracted. \textit{SXT} background file and response matrix file (rmf) provided by the \textit{SXT} instrument team are used. The auxiliary response file (ARF) for the selected source extraction region is obtained using {\it sxtarfmodule}. In the energy spectrum, data is grouped with 20 counts in a single bin. A gain fit correction is applied for \textit{SXT} spectrum using the \textit{gain fit} command to account for instrumental features at low energy values of $1.8$ keV and $2.2 - 2.4$ keV for absorption edges of Si \& Au respectively (see also \citealt{2017JApA...38...29S} and {\it SWIFT-XRT} website\footnote{\url{https://www.swift.ac.uk/analysis/xrt/digest_cal.php\#res}}).
\vspace{-0.5cm}
\subsection{\textit{LAXPC} Data Reduction}
\label{ldr}
\textit{LAXPC} Level-1 data is downloaded from the ISSDC data archive. Level-1 data is processed to Level-2 by using the {\it LAXPC} pipeline software {\it (LaxpcSoft\footnote{\url{https://www.tifr.res.in/~astrosat_laxpc/LaxpcSoft.html}})}. We make use of the data from \textit{LAXPC 20} only for the uniform study of observations during 2016-2020, due to gain instability reported in {\it LAXPC 10} and {\it LAXPC 30} detectors \footnote{\url{http://astrosat-ssc.iucaa.in/}} (see also \citealt{2020MNRAS.tmp.2075B}). Observation specific response and background files generated by the software following \citealt{2017caantialibration} are used. The software routine creates the Good Time Interval (GTI) file consisting of timing information during Earth occultation and South Atlantic Anomaly (SAA). \textit{LAXPC} data is obtained by considering the single events and top layer of the detector unit (see also \citealt{2019MNRAS.487..928S,Katoch2020} (under review)). The analysis and modelling methods of the reduced data are presented in the next section. 
\vspace{-0.6cm}
\section{Analysis and modelling}
\label{AM}
\subsection{Temporal Analysis}
\label{TA}
\hspace{1cm}
Source light curves with a time-bin of 1 day, obtained from \textit{MAXI} observations\footnote{\url{http://maxi.riken.jp/mxondem/}} in the energy range of $2-20$ keV, $2-6$ keV and $6-20$ keV are used to study the long term variance of the sources. The hardness ratio (HR) is obtained by dividing the source flux (in units of photons/cm$^2$/s) of the two light curves (i.e. flux of $6-20/2-6$ keV). We estimate the fractional variance of the light curve for the period of $2016-2020$ using $F_{var}=\left(\sqrt{S^{2}-\overline{\sigma_{err}^{2}}}\right)\Biggm/\overline{x}$ following \citealt{2003MNRAS.345.1271V}. Here, $S^{2}$ is the total variance of the light curve and $\overline{\sigma^{2}_{err}}$ is the error associated with measurement and $\overline{x}$ is the mean count rate of the light curve. \par
We generate the \textit{LAXPC} light curves with 1 sec time-bin to study the short term variability of the sources in the energy range of $3-20$ keV. 
{\it LAXPC} light curves of 50 ms time-bin in the energy range of $3-20$ keV are also considered to study the nature of the PDS. Each light curve is divided into intervals of 8192 time bins and the PDS is constructed for each of these bins. The average of these PDS is obtained and is binned logarithmically by a factor of $1.05$ in the frequency space. The resultant binned PDS are normalized to get the fractional rms spectra (\citealt{1990A&A...230..103B}) and the Poisson noise is subtracted implementing the procedure mentioned by \citealt{2018MNRAS.477.5437A}. The resultant PDS in the frequency range $0.002-10$ Hz is fitted with a {\it powerlaw} distribution function in the form of $P(\nu) = A\nu^{\gamma}$, where $\gamma$ is the power-law index and $A$ is the norm. \par
We also estimate the fractional rms amplitude in the frequency range $0.002-10$ Hz using the rectangle rule integration method given by $rms = \sqrt{(P\times \Delta \nu)} \times 100$ (in \%)\footnote{\url{https://heasarc.gsfc.nasa.gov/docs/xte/recipes/pca_fourier.html}} (see also \citealt{2016MNRAS.460.4403R, 2016MNRAS.462.1834R,2018Ap&SS.363..189D} for details), where $P$ is the power in the units of rms$^{2}$/Hz and $\Delta \nu$ is the frequency in Hz. We follow the above mentioned procedures for both sources and obtained timing results are presented in section \ref{LHV}. 
\begin{figure*}
    \centering
    \includegraphics[width=8cm]{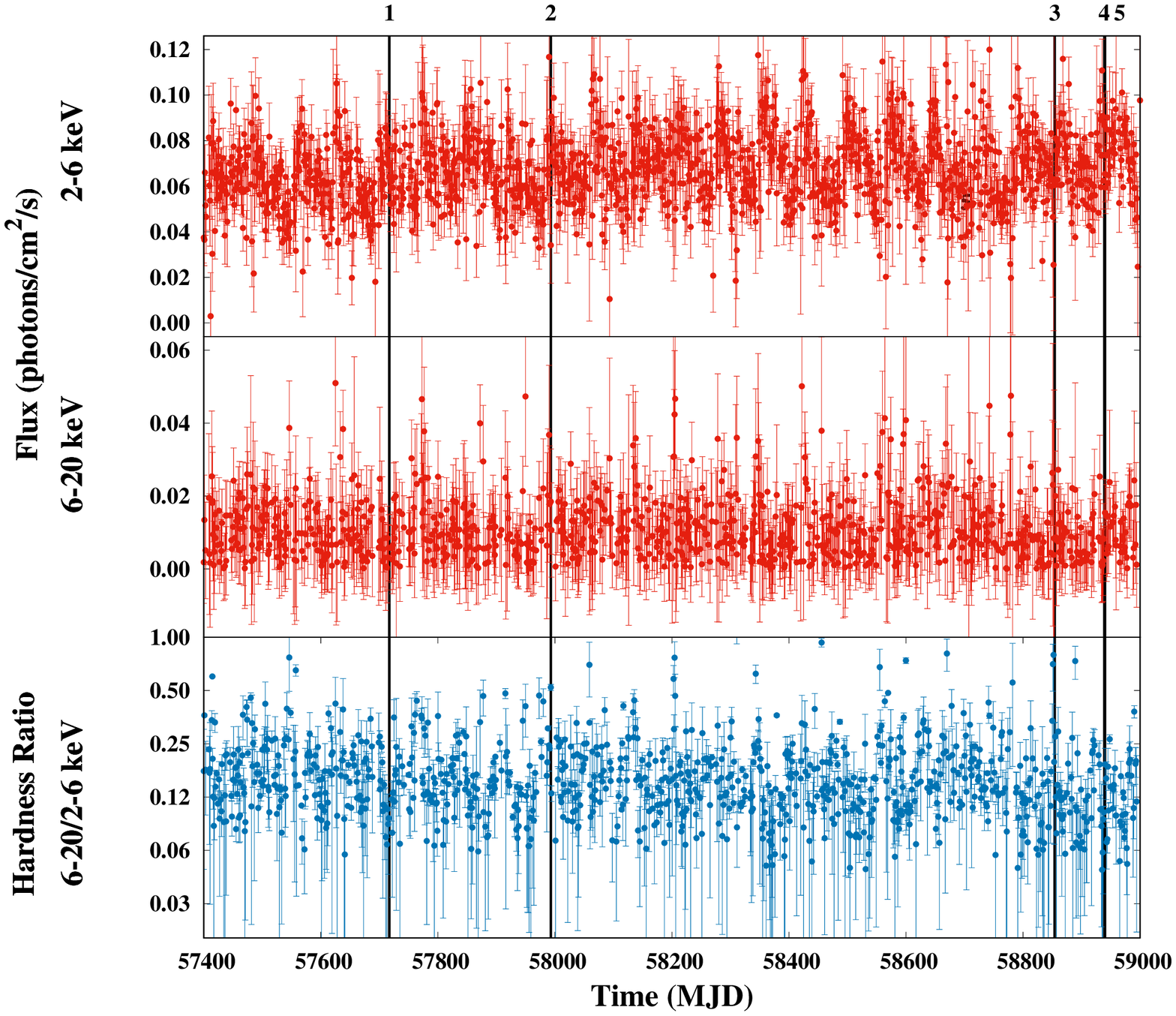} 
    \hspace{1cm}
    \includegraphics[width=8cm]{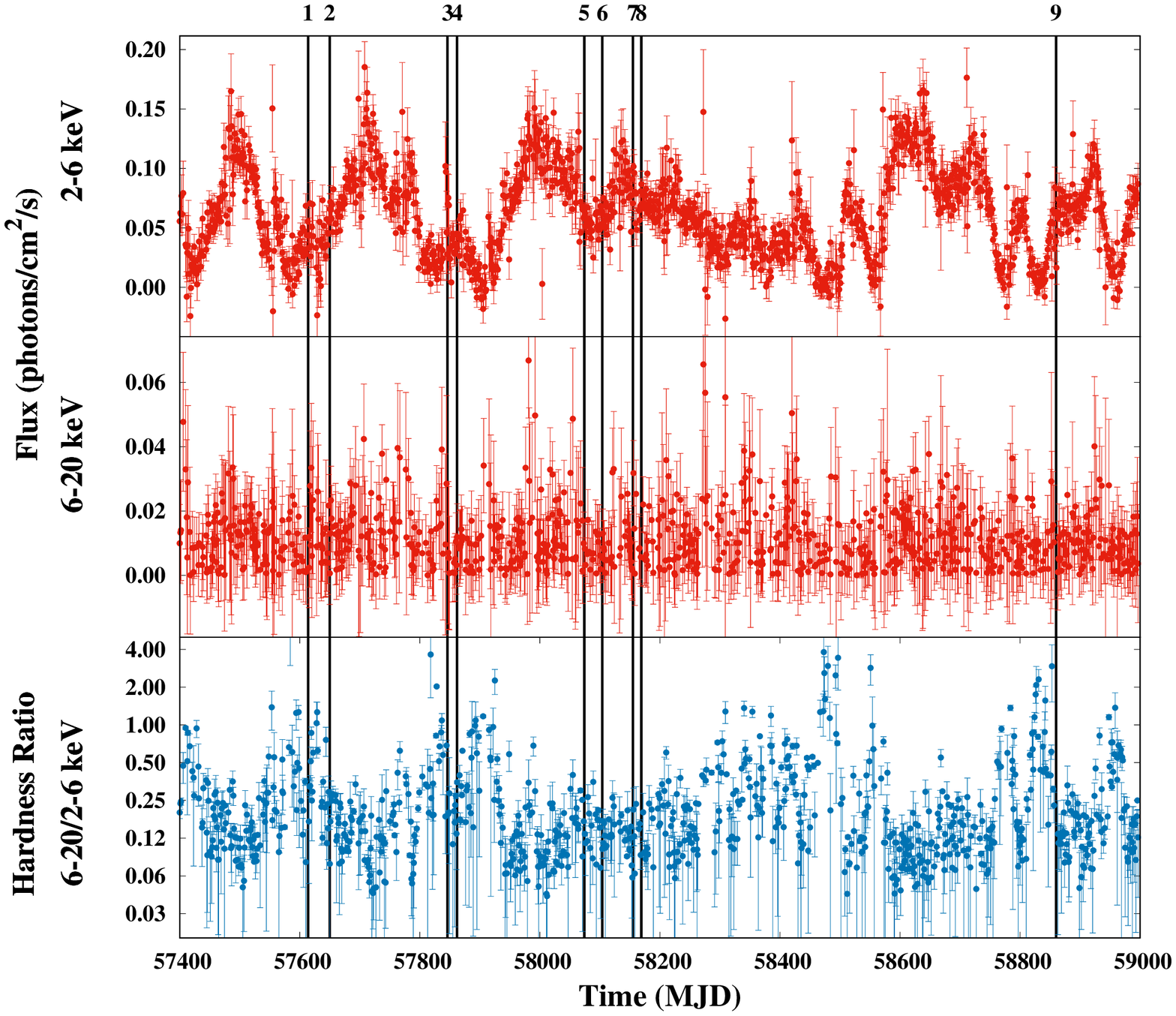}
    \caption{{\it MAXI} light-curve of LMC X-1 (left panel) and LMC X-3 (right panel) during the period of $2016-2020$ in the energy band of $2-6$ keV (top panel) and $6-20$ keV (middle pane). Ratio of the flux in hard energy and soft energy bands ($6-20/2-6$ keV) is plotted in bottom panel. All the \textit{AstroSat} observations considered for these sources have been marked in black vertical lines along-with corresponding Epoch numbers.}
    \label{fig2}
\end{figure*} 
\subsection{Spectral Analysis}
\label{SA}
\hspace{1cm}
As the energy spectra of both sources have no significant flux above $20$ keV, we consider the \textit{SXT} spectrum in the energy range of $0.5-7$ keV and that of \textit{LAXPC} in $4-20$ keV. Broadband energy spectra ($0.5-20$ keV) are analyzed and modeled using \textit{XSpec v12.10}  tool \citep{1996ASPC..101...17A} of {\it HEAsoft v6.26.1}. \par
We use \textit{TBabs} model to represent the hydrogen column density (\textit{$n_H$}) along the line of sight. Since  much of the absorption of X-ray from LMC X-1 and LMC X-3 is due to the metallic abundance in the host galaxy \citep{hanke10}, we consider the LMC abundance following \cite{hanke10} in our calculation of $n_H$. These abundance values are adapted by using the \textit{abund file} command in \textit{XSpec} during the spectral fit. Initially, we model the broadband spectral data using a single \textit{powerlaw} along with an absorption model by keeping the parameters of \textit{powerlaw} in both data set independent of each other. It is observed that the unfolded spectrum with respect to \textit{powerlaw} along-with the ratio of data by \textit{powerlaw} model shows instrumental smearing \citep{1999MNRAS.309..113V,2002MNRAS.329L...1B,2002MNRAS.335L...1F,2003MNRAS.340L..28F} as well as the need for disc component in the lower energy range. Then, in order to study the spectral properties of the source, we model the broadband ($0.5-20$ keV) energy spectrum of LMC X-1 for all the epochs with the phenomenological model \textit{TBabs(diskbb+powerlaw)}. This model combination is  referred as Model-1. An overall systematic error of $1\%$ is incorporated into all the fits to account for uncertainty in the response matrix.  For LMC X-3, energy spectra  of all the epochs are found to be very soft with significant data till $10$ keV. Therefore, we use Model-1 for spectral modeling of LMC X-3 without \textit{powerlaw} component for all observations except Epochs 1 and 4. However, a \textit{smedge} is included to account for the reflection seen above $7$ keV for most of the epochs. We also made an attempt to model the spectra with physical model \textit{nthcomp} \citep{1999MNRAS.309..561Z} of \textit{XSpec} but the fit did not give meaningful physical parameter values and $\chi^2_{red}$ obtained was $>2$. From the broadband energy spectra, we calculate the unabsorbed bolometric source flux (in $0.1-100$ keV) as well contribution of disc flux (in $0.5 - 20$ keV) over the total unabsorbed flux using the \textit{cflux} model. The hardness ratio (HR) is calculated by estimating the ratio of flux in the energy range $6 - 20$ keV and $3 - 6$ keV. Errors for all the parameters are obtained at 90\% confidence interval, and relevant error propagation formulae are also implemented based on \citealt{2003drea.book.....B}. 
\par
Further, we model the  broadband energy spectra using several relativistic accretion disc models in order to constrain the physical parameters of the BHs. In this regard, we consider the \textit{kerrbb} model \citep{2005ApJS..157..335L}, which assumes a thin, relativistic accretion disc around a Kerr black holes. Therefore, we model the broadband energy spectra using the continuum model combination of {\it kerrbb} and {\it simpl} \citep{2009PASP..121.1279S} in order to constrain the mass {\it ($M_{BH}$)}, spin {\it (a)} and mass accretion rate {\it ($\dot{M}$)} of the system. We refer to the model combination: {\it TBabs(kerrbb*simpl)} as Model-2 hereafter. While {\it kerrbb} takes into account of the relativistic effect on soft X-ray from the accretion disc, {\it simpl} model considers the Comptonization process of the disc photons.  While using Model-2 for both sources, we consider the known values of inclination angle and distance (see section \ref{Intro}). The detailed results obtained from timing and spectral modeling are presented in section \ref{res}.

\section{Results}
\label{res}
\subsection{Temporal Properties}
\label{LHV}

In this section, we present the  temporal properties and lightcurve variability of LMC X-1 and LMC X-3 using \textit{LAXPC} data. Along with this, we present the long-term light curve variability of the sources as observed by \textit{MAXI} for a period of $\sim4.5$ years (MJD 57400 - MJD 59000) within the duration of \textit{AstroSat} observations.\\
In the left side of Figure \ref{fig2}, we show the variability of the light curve in $2-6$ keV (top panel), $6-20$ keV (middle panel) and HR (bottom panel) of LMC X-1. In the right side of Figure \ref{fig2}, we plot the same for LMC X-3. It is observed that LMC X-1 does not show significant variability and the average HR of this source is $\sim0.2$. The fractional variance $F_{var}$ of the light curve in the energy range $2-20$ keV, is calculated using the method mentioned in section \ref{TA} is found to be $24.9\%$ for LMC X-1. On the other hand, LMC X-3 shows significant intensity variation with HR in the range $0.01-2.0$. The $F_{var}$ calculated for LMC X-3 considering the entire time period is found to be $52.9\%$. The light curve periodicity during the initial $700$ days is $\sim100 - 200$ days and beyond that, the source has a random variability pattern (see top right panel of Figure \ref{fig2} for LMC X-3). \\
Short term variability of the source is studied using \textit{LAXPC} light curve. It is found that the average source intensity for LMC X-1 varies in the range of $39 − 52$ counts/sec ($3 - 20$ keV) and HR in $0.10-0.33$ between different epochs. In the case of LMC X-3, the source count rate in different observations varies between $50 − 240$ counts/sec ($3 - 20$ keV) and HR is in the range $0.07 − 0.21$. The fractional variance of light curve is estimated to be varying in the range of $7.41\%-15.89\%$ and $9.7\%-23.9\%$ during the different epochs of LMC X-1 and LMC X-3 respectively.

\begin{figure*}
\includegraphics[width=6cm,height=7cm]{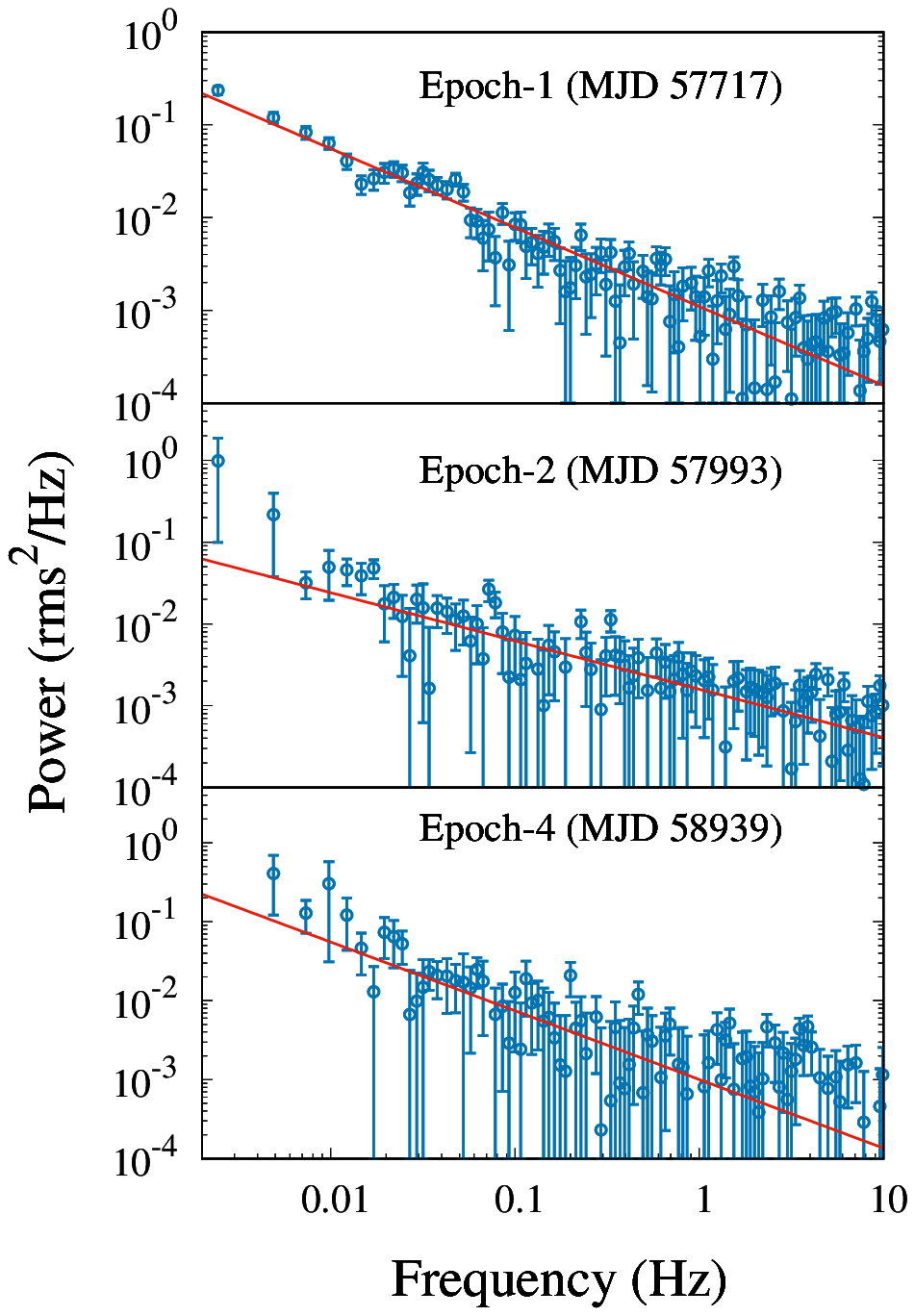} 
\includegraphics[width=9cm,height=6cm]{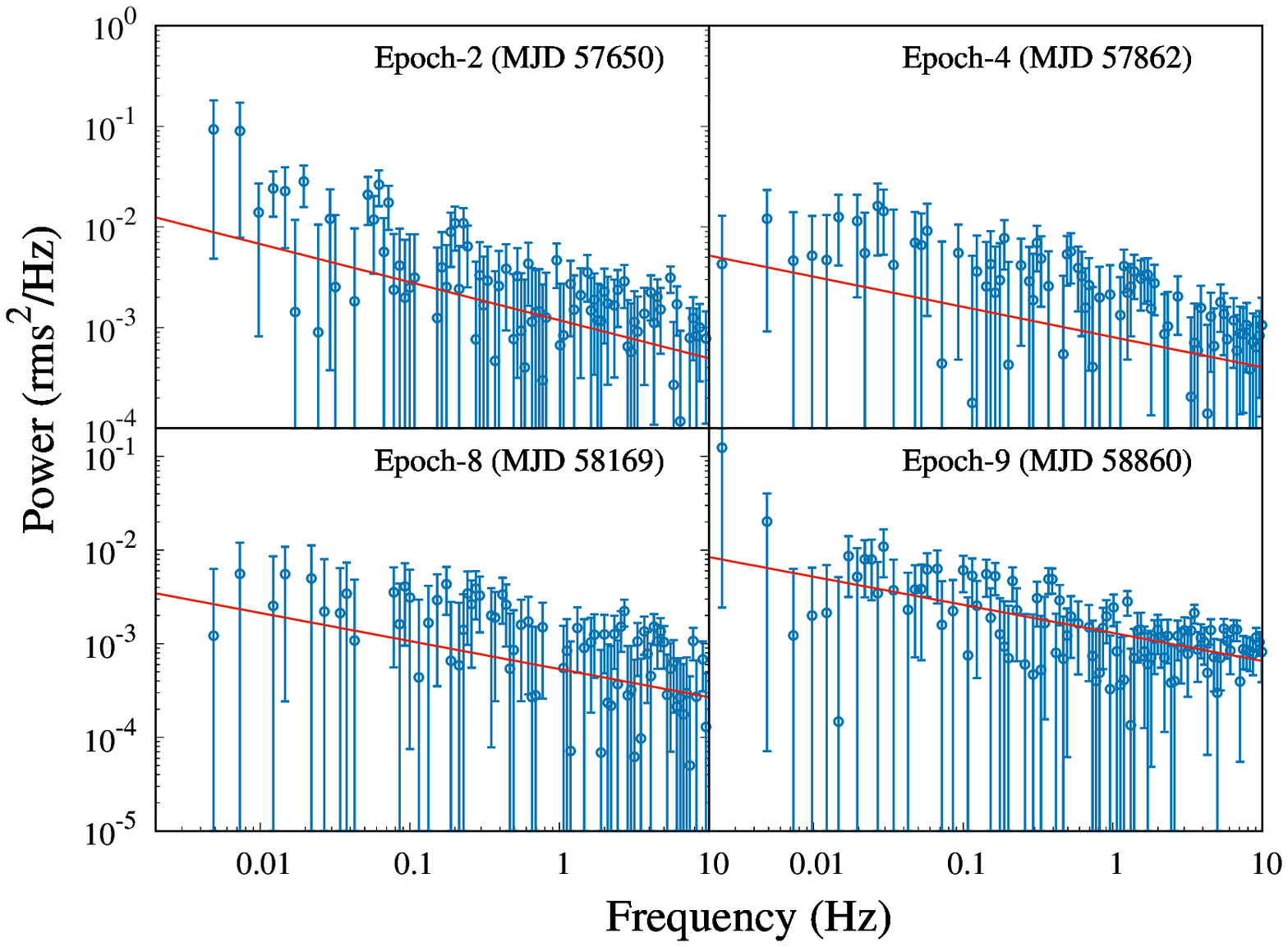}
\caption{Power spectra of both sources (LMC X-1 in left panel and LMC X-3 in right panel) during different observations in the frequency range of $0.002-10$ Hz. The PDS are modeled with different {\it powerlaw} components whose values are quoted in Table \ref{table2}. See text for details.}
\label{fig3}
\end{figure*}

As mentioned in section \ref{TA}, we obtain the power spectrum for each observation of both sources in the frequency range of $0.002-10$ Hz using \textit{LAXPC} lightcurve with 50 msec binning. We model the PDS using \textit{powerlaw} model for all observations of LMC X-1 (shown for epochs 1, 2 and 4 in the left panel of Figure \ref{fig3}). No narrow frequency signatures are seen in the PDS. The fractional rms amplitude in the frequency range $0.002-10$ Hz estimated to be varying from $9.86\%$ to $16.7\%$ during the different epochs. Similarly, PDS for all observations of LMC X-3 are fitted using {\it powerlaw} model. Fitted PDS corresponding to four epochs are shown in the right panel of Figure \ref{fig3}. None of the PDS show any narrow frequency signature. The fractional rms amplitude of the PDS is in the range of $7.16-10.9\%$. In Table \ref{table2}, we present estimated parameters from the temporal study for both sources.

\begin{table*}
\caption{Details of results obtained for temporal analysis of the sources LMC X-1 and LMC X-3. The PDS corresponds to frequency range of $0.002-10$ Hz in the energy band $3-20$ keV. {\it powerlaw} model is used to fit the PDS with power-law index $\gamma$ and norm $n_{PL}$. Fractional rms amplitude value calculated is given in \%. We also quote the fractional variance $F_{var}$, estimated for variability of \textit{LAXPC} light curve which is discussed in section \ref{LHV}. Errors are calculated with $90\%$ confidence for all the parameters. }
\label{table2}	
\begin{tabular}{cccccc}
\hline
\multicolumn{6}{c}{LMC X-1}                                                                                                                                                                                                                   \\ \hline 
Epoch & $\gamma$             & \begin{tabular}[c]{@{}c@{}}$n_{PL}^{\dagger}$ \\ (in units of $10^{-3}$)\end{tabular}                              & fractional rms (\%) & $\chi^{2}/dof$ & \begin{tabular}[c]{@{}c@{}}$F_{var}$\\ (\%)\end{tabular} \\ \hline
1     & $-0.85 \pm 0.04$  & 1.10 & 10.14           & $125.9/109=1.15$             & 11.9                                                     \\
2     & $-0.59 \pm 0.09$  & 1.59       & 12.1           & $122.1/109=1.12$             & 13.14                                                     \\
      3     & $-0.65 \pm 0.04$  & 1.15        & 8.99           & $88.30/109=0.81$             & 7.41                                                     \\
      4     & $-0.87^{+0.16}_{-0.22}$ & 1.01       & 16.7           & $105.8/109=0.97$             & 15.89                                                     \\
      5    & $-0.83 \pm 0.03$  & 1.23      & 9.86           & $117.1/109=1.09$            & 11.18                                                     \\
       \hline
\multicolumn{6}{c}{LMC X-3}                                                                                                                                                                                                                   \\ \hline
1     & $-0.20^{*}$ & 1.15   & 8.37            & $141.7/108=1.31$             & 9.7                                                      \\
2     & $-0.37^ \pm 0.09$ & 1.18                                         & 10.8           & $115.5/109=1.08$             & 16.1                                                     \\
3     & $-0.72^{+0.25}_{-0.16}$ & 0.83                                       & 10.9           & $122.0/108=1.13$             & 19.0                                                     \\
4     & $-0.3^{*}$ & 1.20                                           & 8.8           & $115.2/109=1.06$             & 23.9                                                     \\
5     & $-0.25^{+0.14}_{-0.20}$ & 1.12                                         & 9.9            & $117.2/108=1.08$             & 14.7                                                     \\
6     & $-0.2^{*}$ & 0.60                                        & 7.16            & $99.4/109=0.91$             & 12.3                                                     \\
7     & $-0.3^{*}$ & 0.66                                       & 7.6            & $113.9/109=1.04$             & 11.5                                                     \\
8     &  $-0.3^{*}$ & 0.59                                        & 7.09            & $127.4/109=1.17$             & 17.4                                                     \\   
9     & $-0.14^{+0.09}_{-0.08}$  & 1.00                                       & 10.3            & $110.5/109=1.01$             & 11.9                                                     \\
\hline
\end{tabular}
\begin{tablenotes}

\small

\item{$^{*}$ Frozen}
\item{$^{\dagger}$ The error values are trivial and hence not quoted.} 
\end{tablenotes}
\end{table*}
\subsection{Spectral Properties}
\subsubsection{LMC X-1}
\label{SPX1}
\begin{figure*}
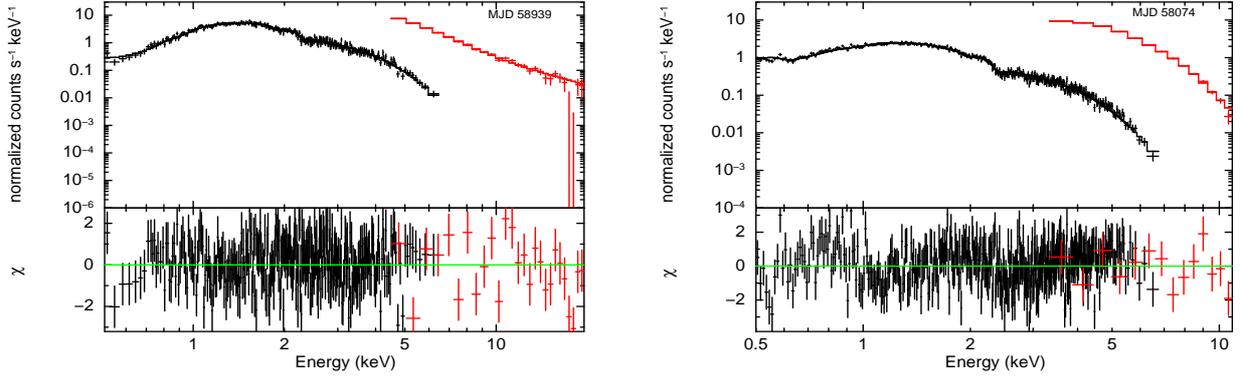

\includegraphics[height=8cm,width=5cm,angle=-90]{lmcx1_pheno.eps}
\hskip 0.2in
\includegraphics[height=8cm,width=5cm,angle=-90]{lmcx3_pheno.eps}
\caption{Broadband spectral fitting performed for Epoch-4 observation of LMC X-1 (left) in $0.5-20$ keV and Epoch-5 observation of LMC X-3 (right) in $0.5 - 10$ keV. Combined \textit{SXT} and \textit{LAXPC} spectrum is modelled with Model-1 for LMC X-1 and \textit{diskbb} along with \textit{smedge} for LMC X-3.}
\label{fig4}
\end{figure*}

\textit{AstroSat} has observed the source LMC X-1 during five epochs (see Table \ref{table1}) with a time gap of several months. From the spectral fit, we get the $n_{H}$ value of $1.22 \pm 0.02 - 1.48 \pm 0.04 \times 10^{22}$ during different epochs. These values are greater than that obtained using only the Galactic abundance \citep{2000ApJ...542..914W} which is $\sim0.6 \times 10^{22}$ atoms $cm^{-2}$. We note that the Epoch-1 spectrum requires an additional cut-off at $6.29^{+0.08}_{-0.09}$ keV which is taken into account with a {\it highecut} model ($\chi^2_{red}$ improves from 1.94 to 1.32). During the different epochs, the value of disc temperature ($T_{in}$) varies from $0.86\pm{0.01} - 0.98\pm{0.01}$ keV, while the photon index ($\Gamma$) is in the range $2.46\pm{0.04} - 3.29\pm{0.04}$.  
The bolometric luminosity calculated ($0.1-100$ keV)  is in the range of $8.4\times10^{37}$ erg/s ($\equiv 0.07$ $L_{Edd}$) to $1.17\times10^{38}$ erg/s ($\equiv 0.10$ $L_{Edd}$). Total flux is found to be contributed mainly from the disc with $79$\% - $94$\% contribution. Also, the value of the hardness ratio is observed to vary from $0.16$ to $0.34$. We do not observe any feature of Fe line emission in any of the epochs, but a reflection edge exists $> 7$ keV during Epochs 1 and 2. The folded spectrum of Epoch-4 of LMC X-1 is shown in the left plot of Figure \ref{fig4}. In Table \ref{table3}, we summarize the values obtained for the parameters using Model-1. 

The observed inner disc radius of the accretion disc $r_{in}$ is calculated using the normalization ($N_{diskbb}$) value of \textit{diskbb}. $N_{diskbb}$ and disc radius are related as $N_{diskbb}=(r_{in}/D_{10})^{2} \times \cos \theta$, where $r_{in}$ is the observed inner disc radius, $D_{10}$ is the distance to the source in 10 kpc and $\theta$ is the inclination angle of the disc. However, since $r_{in}$ is the apparent radius we estimate the `true' radius following \citealt{1998PASJ...50..667K}. `True' radius is given as $R_{in}=\xi.\kappa^{2}.r_{in}$ where $\kappa$ is taken to be 1.55 (see section \ref{Mna} for details) and $\xi=0.412$ as given by \citealt{1998PASJ...50..667K}. We obtain the value of `true' radius $R_{in}$ to be varying from $29.65\pm{0.71}$ km to $39.17\pm{0.71}$ km during the different epochs of LMC X-1.   

\begin{table*}
\caption{Spectral parameter values obtained from the best-fit of Model-1 for both LMC X-1 and LMC X-3. $T_{in}$ (in units of keV) corresponds to disc temperature, $N_{diskbb}$ is disc norm. The parameters of \textit{powerlaw} are denoted by the photon index $\Gamma$ and norm $N_{PL}$. The value of $edgeE$ is the energy component from the \textit{smedge} model. $L_{bol}$ is the bolometric luminosity for the energy range $0.1-100$ keV in units of $10^{38}$ erg/s estimated by incorporating the known values of mass quoted in section \ref{Intro}. $f_{disc}$ and $f_{total}$ are unabsorbed disc flux and total flux calculated in the energy range $0.5-20$ keV. $HR$ corresponds to the hardness ratio estimated for ratio of flux in $6-20$ keV and $3-6$ keV. Error for all the parameters are calculated with $90\%$ confidence, and by also considering the error propagation method wherever appropriate.}
\label{table3}
\begin{tabular}{cccccccccc}
\hline
\multicolumn{9}{c}{LMC X-1}                                                                                                                                                                                                                                                                                                                                                                                                      \\ \hline
Epoch & \begin{tabular}[c]{@{}c@{}}$T_{in}$ \\ (keV)\end{tabular} & $N_{diskbb}$            &  $\Gamma$               & $N_{PL}$            & \begin{tabular}[c]{@{}c@{}}$edgeE$\\ (keV)\end{tabular} & \begin{tabular}[c]{@{}c@{}}$L_{bol}$\\  ($\times 10^{38}$ erg/s)\end{tabular} & \begin{tabular}[c]{@{}c@{}}$f_{disc}/f_{total}$\\ (\%)\end{tabular} & $HR$ & $\chi^{2}/dof$  \\ \hline
1$^{*}$     & $0.86 \pm 0.01$                                  & $53.41^{+3.15}_{-3.09}$ &  $2.63 \pm 0.02$                      & $0.10^{\ddagger}$                      & $7.54^{+0.87}_{-0.70}$                                                     & $0.94$                                               & $84$                                                 &        $0.21 \pm 0.02$          & $700.39/529=1.32$          \\
2     & $0.90^{+0.02}_{-0.01}$                                     & $46.17^{+4.86}_{-4.35}$    & $2.46 \pm 0.04$ & $0.16 \pm 0.01$ & $8.28^{+0.36}_{-0.35}$                                                     & $1.17$                                               & $82$                                                         &$0.34 \pm 0.02$         & $582.95/417=1.39$            \\
3     & $0.96 \pm 0.01$                                  & $35.02^{+1.77}_{-2.23}$                                                        & $3.29 \pm 0.04$                      & $0.10^{\ddagger}$                     & -                                                     & $0.84$                                               & $86$                                                                &$0.16 \pm 0.02$  & $670.77/506=1.32$           \\ 
4     & $0.98 \pm 0.02$                                  & $30.96^{+1.77}_{-0.91}$  & $3.03^{+0.13}_{-0.16}$                      & $0.09^{\ddagger}$                      & -                                                     & $1.02$                                               & $79$                                                                & $0.24 \pm 0.03$ & $393.08/338=1.16$            \\
      5     & $0.98 \pm 0.01$                                  & $30.74^{+0.29}_{-0.17}$                                                        & $3.19^{+0.02}_{-0.03}$                   & $0.15 \pm 0.01$                    & -                                                     & $0.95$                                               & $94$                                                       & $0.19 \pm 0.02$          & $540.27/480=1.12$           \\
\hline
\multicolumn{9}{c}{LMC X-3}                                                                                                                                                                                                                                                                                                                                                                                                      \\ \hline
1     & $1.08 \pm 0.01$                                    & $16.18^{+0.97}_{-0.90}$                                                                & $3.60 \pm 0.09$         & $0.01^{\ddagger}$     & -                                                     & $0.65$                                               & $97$                                                               & $0.19 \pm 0.02$ & $466.08/403=1.16$        \\
2     & $1.10 \pm 0.01$                                    & $18.24^{+0.57}_{-0.55}$                                                        & -                      & -                      & $8.29^{+0.30}_{-0.29}$                                & $0.76$                                               & $97$                                                      & $0.17 \pm 0.02$            & $503.81/393=1.28$         \\
3     & $0.99 \pm 0.01$                                    & $21.04^{+1.24}_{-0.60}$                                                          & -                      & -                      & $8.16 \pm 0.03$                                & $0.52$                                               & $96$                                                        & $0.14 \pm 0.02$          & $430.30/417=1.03$           \\
4     & $1.00 \pm 0.02$                                    & $18.19^{+2.25}_{-2.95}$                                                                & $2.78^{+0.21}_{-0.18}$         & $0.01^{\ddagger}$     & -                                                     & $0.53$                                             & $96$                                                               & $0.17 \pm 0.05$  & $326.11/318=1.03$         \\
5     & $1.09 \pm 0.01$                                     & $18.34^{+0.56}_{-0.55}$                                                          & -                      & -                      & $7.90^{+0.35}_{-0.36}$                                & $0.74$                                               & $98$                                                             &$0.17 \pm 0.01$       & $529.48/425=1.24$          \\
6     & $1.11 \pm 0.01$                                            & $20.81^{+1.49}_{-1.71}$                                                                & -                      & -                      & $8.20^{+1.14}_{-0.69}$                                        & $0.93$                                               & $99$                                                                 &$0.18 \pm 0.04$ & $460.08/452=1.02$           \\
7     & $1.13 \pm 0.01$                                    & $20.16^{+0.52}_{-0.77}$                                                          & -                      & -                      & $7.99^{+0.10}_{-0.13}$                                  & $0.98$                                               & $98$                                                        & $0.18 \pm 0.01$        & $595.93/453=1.31$          \\
8     & $1.13^{+0.01}_{-0.02}$                                    & $21.35^{+1.30}_{-1.72}$                                                                & -                      & -                      & $7.57^{+0.44}_{-0.58}$                                        & $1.02$                                               & $96$                                                               & $0.18 \pm 0.01$ & $535.65/447=1.20$             \\
      9     & $1.03 \pm  0.01$                                    & $17.18^{+0.97}_{-0.86}$                                                          & -                      & -                      & $6.79 \pm 0.07$                                  & $0.53$                                               & $100^{\dagger}$                                                    & $0.10 \pm 0.03$            & $669.09/480=1.39$            \\
\hline
\end{tabular}
\begin{tablenotes}
\item {$^{*}$ - includes a high energy cut off at $6.29$ keV (see section \ref{SPX1} for details).}
\item {$^{\dagger}$ - only disc contribution.}
\item {$^{\ddagger}$ - error is insignificant.}
\end{tablenotes}
\end{table*}

\subsubsection{LMC X-3}
\label{SPX3}
LMC X-3 has been observed by \textit{AstroSat} during nine epochs with a time gap of several months (see Table \ref{table1}). 
The $n_H$ value obtained during different epochs is in the range of $0.03 \pm 0.01 - 0.07 \pm 0.02 \times 10^{22}$ atoms cm$^{-2}$. Its average value is slightly higher than that obtained by considering only the Galactic abundance ($\sim 0.04 \times 10^{22}$ atoms $cm^{-2}$). During the different epochs, the value of disc temperature is $\sim1$ keV while the norm varies from $16.18^{+0.97}_{-0.90}$ to $21.35^{+1.30}_{-1.72}$. No Fe line features are observed in any of the spectra. Unabsorbed bolometric luminosity ($0.1-100$ keV) is found to vary between $5.10\times10^{37}-1.02\times10^{38}$erg/s ($\equiv 0.07 - 0.13$ $L_{Edd}$).
The disc flux contribution and hardness ratio calculated are $>96\%$ and $0.10 - 0.19$ respectively throughout the different observational epochs. Modified inner disc radius $R_{in}$ is found to be in the range $32.40\pm{0.95} - 37.21\pm{0.95}$ km, considering $\kappa = 1.7$.  We plot the folded spectrum of Epoch-5 data of LMC X-3 in Figure \ref{fig4}. The spectral parameters of best-fit values are mentioned in Table \ref{table3}. \par

\subsection{Constraining the physical parameters of the black hole}
\label{Mna}
We understand from sections \ref{SPX1} and \ref{SPX3} that both sources have a soft energy spectrum dominated by the contribution from the disc component flux. Hence we attempt to estimate the source mass and spin using the broadband continuum fitting method. For this purpose, we model the spectra with Model-2.\par
The important model parameters obtained are $\dot{M}$, $a$, $M_{BH}$ in units of $M_{\odot}$, photon index ($\Gamma$) and scattering fraction ($FracScat$). We estimate $\dot{M}$ of LMC X-1 to be of $1.24^{+0.10}_{-0.11} - 2.16^{+0.39}_{-0.17} \times10^{18}$ g/s. We find that this value of accretion rate is $0.29 - 0.51$ of the Eddington rate ($\dot{M_{Edd}}$). \citealt{2009ApJ...701.1076G} estimated the spectral hardening factor for this source to be $1.55$ using {\it RXTE} observations during which the source existed in a thermal dominated state. Since all the {\it AstroSat} observations considered in this paper also have thermal disc dominated spectra with almost constant luminosity, we choose the same value for the hardening factor. To confirm the consistency of this value, we fit the spectral data by keeping hardening factor as a free parameter, which resulted in a value of $1.54-1.56$ for different epochs. Hence, we consider the hardening factor to be 1.55 for all the epochs of LMC X-1. We note that the resultant parameters using the {\it simpl} model i.e. $\Gamma$ is $\sim2.61^{+0.15}_{-0.19} - 4.50^{+0.03}_{-1.28}$ and the $FracScat$ is in the range $0.03^{+0.03}_{-0.01} - 0.2^{+0.03}_{-0.02}$ for different epochs of LMC X-1. The $\dot{M}$ of LMC X-3 is in the range $2.02^{+0.09}_{-0.06}$ to $4.03^{+0.17}_{-0.14}\times10^{18}$ g/s which corresponds to $0.14 - 0.29$ of $\dot{M_{Edd}}$. Spectral hardening factor of $1.7$ is considered for these fits of LMC X-3.\par
\begin{figure*}
	\hskip -0.2in
\includegraphics[width=7cm]{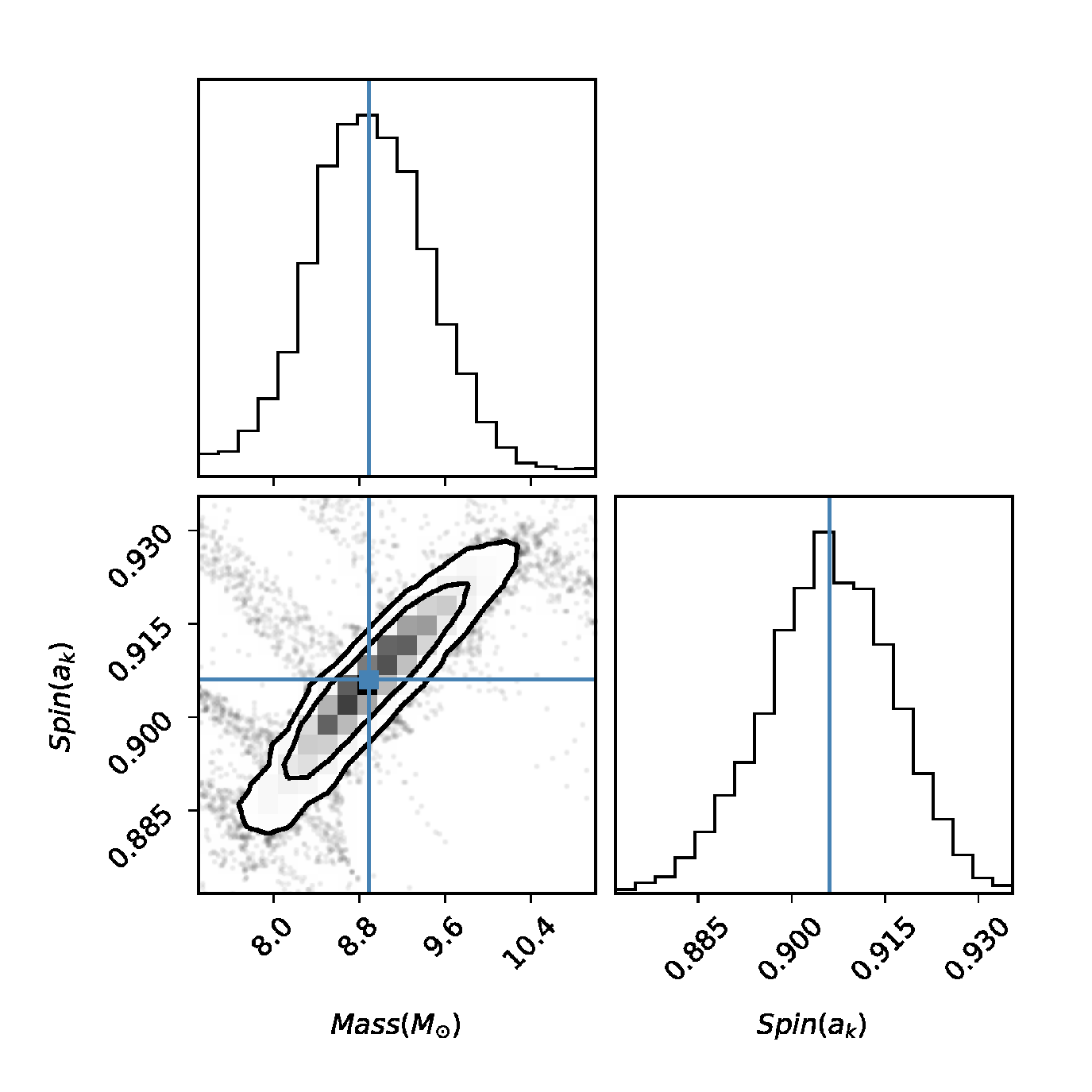}
\includegraphics[width=7cm]{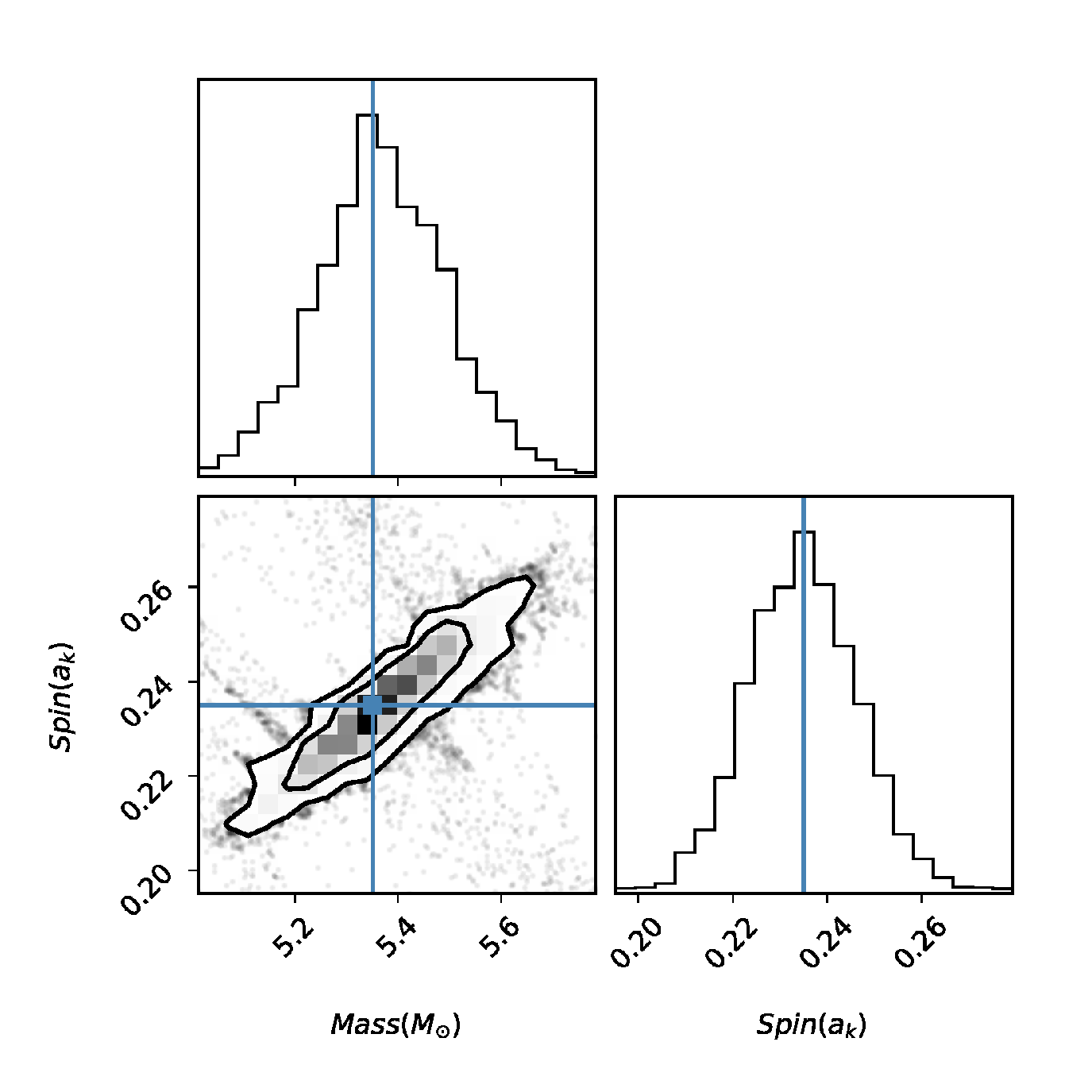}
\caption{Contour plot of mass versus spin plotted for Epoch-4 of LMC X-1 (left plot) and Epoch-7 of LMC X-3 (right plot) adapting MCMC Hammer algorithm. The corner plot shows the contour of mass versus spin with 68 and 90\% confidence. Top panel shows the probability distribution of mass and the right panel shows the distribution of spin. }
\label{fig6}
\end{figure*}

We obtain the source mass for LMC X-1 to be in the range of $7.64^{+0.99}_{-0.25}$ $M_{\odot}$ to $10.00^{+0.52}_{-1.22}$ $M_{\odot}$ during the different epochs. The spin parameter is found to be varying from $0.82^{+0.10}_{-0.02} - 0.92^{+0.03}_{-0.04}$. During the nine observation epochs of LMC X-3, we find the source mass to be varying from  $5.35^{+0.34}_{-0.35}$ $M_{\odot}$ to $6.22^{+0.48}_{-1.74}$ $M_{\odot}$. The value of spin is estimated to be in the range of $0.22 \pm 0.02 - 0.41 \pm 0.02$. The model fitted parameters along-with the values of spin and mass are tabulated in Table \ref{table4}. In order to get a better estimation of error of the mass and spin parameters, we have performed Markov Chain Monte Carlo (MCMC) chain simulation using Goodman-Weare algorithm \citep{2010CAMCS...5...65G}. This is incorporated using \textit{XSpec} where the \textit{walkers} parameter is set to 32, initial chain length is taken to be 15,000 with a burn length of 5,000. In Figure \ref{fig6}, we show the contour plots of mass versus spin values and their probability distributions for the Epoch-4 and Epoch-7 observations of LMC X-1 and LMC X-3 respectively. This is plotted by adapting the MCMC Hammer algorithm \citep{2013PASP..125..306F}. The contours are plotted for the 68\% and 90\% confidence intervals (see also \citealt{2020MNRAS.499.5891S}), showing the range of spin and mass values.  
\begin{table*}
\caption{Best-fit parameters obtained by fitting Model-2 for continuum-fitting of broadband energy spectra. The value of $d$ is frozen to $48.1$ kpc, while $i$ is considered as $36.38\pm{1.92}^{\circ}$ for LMC X-1 and $69.24\pm{0.72}^{\circ}$ for LMC X-3 (see section \ref{Intro}). BH mass $M_{BH}$ is constrained to be in the range of $7.0-12.0 M_{\odot}$ and $5.0-8.0M_{\odot}$ for LMC X-1 and LMC X-3 respectively. Spectral hardening factor of $1.55$ and $1.7$ has been considered for this fitting of LMC X-1 and LMC X-3 respectively. Parameter $\dot{M}$ is the accretion rate in units of $10^{18}$ g/s, $\Gamma$ is the photon index, $FracScat$ is the scattering fraction, $a$ represents the source spin and $M_{BH}$ is the mass in units of $M_{\odot}$. Error of all the parameters are estimated at $90\%$ confidence.}
\label{table4}
\begin{tabular}{ccccccc}
\hline
\multicolumn{7}{c}{LMC X-1}                                                                                                                                                                                                                       \\ \hline
Epoch & $\dot{M}$  & $\Gamma^*$ & $FracScat^*$ & $a$ & $M_{BH}$ & $\chi^{2}/dof$ \\
 & $\times10^{18}$(g/s) & & & &  ($M_{\odot}$) & \\
\hline
1                     & \multicolumn{1}{c}{$1.60^{+0.01}_{-0.16}$} & \multicolumn{1}{c}{$4.46^{+0.24}_{-0.31}$}& \multicolumn{1}{c}{$0.20^{+0.03}_{-0.02}$} & \multicolumn{1}{c}{$0.86^{+0.08}_{-0.01}$} & \multicolumn{1}{c}{$9.39^{+0.50}_{-0.46}$}                                      & \multicolumn{1}{c}{$744.50/529=1.40$}             \\
2                    & \multicolumn{1}{c}{$2.16^{+0.39}_{-0.17}$}  & \multicolumn{1}{c}{$2.61^{+0.15}_{-0.19}$}& \multicolumn{1}{c}{$0.13^{+0.01}_{-0.02}$} & \multicolumn{1}{c}{$0.82^{+0.10}_{-0.02}$} & \multicolumn{1}{c}{$10.00^{+0.52}_{-1.22}$}                                          & \multicolumn{1}{c}{$682.18/417=1.50$}             \\
3                     & \multicolumn{1}{c}{$1.24^{+0.10}_{-0.11}$}  & \multicolumn{1}{c}{$3.04^{+1.13}_{-0.78}$}& \multicolumn{1}{c}{$0.03^{+0.03}_{-0.01}$}& \multicolumn{1}{c}{$0.90 \pm 0.03$} & \multicolumn{1}{c}{$7.64^{+0.99}_{-0.25}$}                                          &  \multicolumn{1}{c}{$631.94/510=1.24$}             \\
4                     & \multicolumn{1}{c}{$1.48^{+0.24}_{-0.15}$}   &\multicolumn{1}{c}{$3.31^{+0.57}_{-0.36}$}&\multicolumn{1}{c}{$0.09^{+0.05}_{-0.02}$}  & \multicolumn{1}{c}{$0.91^{+0.03}_{-0.07}$} & \multicolumn{1}{c}{$8.92^{+1.22}_{-1.13}$}                                          & \multicolumn{1}{c}{$397.53/338=1.18$}             \\
5                     & \multicolumn{1}{c}{$1.29^{+0.12}_{-0.13}$}  & \multicolumn{1}{c}{$4.50^{+0.03}_{-1.28}$}& \multicolumn{1}{c}{$0.09^{+0.04}_{-0.05}$} & \multicolumn{1}{c}{$0.92^{+0.03}_{-0.04}$} & \multicolumn{1}{c}{$7.99^{+0.76}_{-0.44}$}                                          & \multicolumn{1}{c}{$545.63/490=1.11$}             \\
\hline
\multicolumn{7}{c}{LMC X-3}                                                                                                                                                                                                                       \\ \hline
1                     & \multicolumn{1}{c}{$2.88^{+0.07}_{-0.06}$}    & \multicolumn{1}{c}{-} & \multicolumn{1}{c}{-} & \multicolumn{1}{c}{$0.30^{+0.01}_{-0.03}$} & \multicolumn{1}{c}{$5.89^{+0.17}_{-0.26}$} & \multicolumn{1}{c}{$494.27/403=1.22$}             \\
2                     & \multicolumn{1}{c}{$3.39 \pm 0.11$}   &\multicolumn{1}{c}{-} & \multicolumn{1}{c}{-}  & \multicolumn{1}{c}{$0.30 \pm 0.02$} & \multicolumn{1}{c}{$5.82^{+0.22}_{-0.18}$} & \multicolumn{1}{c}{$472.38/394=1.20$}             \\
3                     & \multicolumn{1}{c}{$2.66^{+0.07}_{-0.06}$}   & \multicolumn{1}{c}{-} & \multicolumn{1}{c}{-} & \multicolumn{1}{c}{$0.30^{+0.03}_{-0.01}$} & \multicolumn{1}{c}{$5.70^{+0.15}_{-0.22}$} & \multicolumn{1}{c}{$424.05/419=1.01$}             \\
4                                                             & \multicolumn{1}{c}{$2.51^{+0.89}_{-0.13}$}   & \multicolumn{1}{c}{-} & \multicolumn{1}{c}{-}   & \multicolumn{1}{c}{$0.38^{+0.01}_{-0.04}$} & \multicolumn{1}{c}{$6.22^{+0.48}_{-1.74}$}                                        & \multicolumn{1}{c}{$316.21/314=1.01$}             \\
5                    & \multicolumn{1}{c}{$2.97^{+0.11}_{-0.05}$}   & \multicolumn{1}{c}{-} & \multicolumn{1}{c}{-}    & \multicolumn{1}{c}{$0.39^{+0.12}_{-0.01}$} & \multicolumn{1}{c}{$6.22^{+0.18}_{-0.17}$}                                        & \multicolumn{1}{c}{$572.34/423=1.35$}             \\
6                                                            & \multicolumn{1}{c}{$4.03^{+0.17}_{-0.14}$}  &\multicolumn{1}{c}{-} &  \multicolumn{1}{c}{-}   & \multicolumn{1}{c}{$0.22 \pm 0.03$}   & \multicolumn{1}{c}{$5.35^{+0.34}_{-0.35}$}                                        & \multicolumn{1}{c}{$567.15/452=1.25$}             \\
7                     & \multicolumn{1}{c}{$3.98^{+0.04}_{-0.02}$}   &\multicolumn{1}{c}{-} &\multicolumn{1}{c}{-}    & \multicolumn{1}{c}{$0.23^{+0.02}_{-0.01}$} & \multicolumn{1}{c}{$5.35^{+0.20}_{-0.17}$}                                         & \multicolumn{1}{c}{$536.32/453=1.18$}             \\
8                    & \multicolumn{1}{c}{$3.97^{+0.16}_{-0.14}$}    &\multicolumn{1}{c}{-}&\multicolumn{1}{c}{-}    & \multicolumn{1}{c}{$0.29 \pm 0.02$}   & \multicolumn{1}{c}{$5.29^{+0.28}_{-0.16}$}& \multicolumn{1}{c}{$600.85/447=1.34$}             \\
9                     & \multicolumn{1}{c}{$2.02^{+0.09}_{-0.06}$}   &\multicolumn{1}{c}{-} & \multicolumn{1}{c}{-}    & \multicolumn{1}{c}{$0.41 \pm 0.02$}   & \multicolumn{1}{c}{$5.50^{+0.10}_{-0.07}$}                                        & \multicolumn{1}{c}{$663.31/480=1.38$}             \\
\hline
\end{tabular}
\begin{tablenotes}

\small

\item{$^{*}$ \textit{simpl} model required only for those epochs with significant data $>10$ keV (see text for details).}
\item{$^{\dagger}$ Frozen}
\item{$^{\ddagger}$ Upper limit}
\end{tablenotes}

\end{table*}

\section{Discussion and conclusions}
\label{DC}
\hspace{1cm}
In this paper, we have studied the broadband spectral and temporal variability of the BH sources LMC X-1 and LMC X-3 using \textit{AstroSat} archival and legacy data. For LMC X-1 we study the source characteristics during five different observation epochs and for LMC X-3 nine observations have been looked into; all spanning over the {\it AstroSat} era of  $\sim4.5$ years. \par
We studied the evolution of source light curve and hardness ratio over long term and short term duration using {\it MAXI} and {\it AstroSat-LAXPC} observations. The fractional variance has been estimated to be around $24.9\%$ for LMC X-1, from the long term {\it MAXI} light curve. Whereas in case of LMC X-3, the long term light curve exhibits a high value of variance $\sim 53\%$ over $4.5$ years. Applying the same procedure as discussed in section \ref{TA} to the entire {\it RXTE-ASM} light curve\footnote{\url{http://xte.mit.edu/asmlc/One-Day.html}} for $\sim15$ years in $3 - 12$ keV band, we obtain the fractional variance as 34\% and 76\% for LMC X-1 and LMC X-3 respectively. On comparing this with the results we obtained using \textit{MAXI} data, it agrees that variability of LMC X-1 on large time scale is less whereas that of LMC X-3 is very high.\par
We note that the short term {\it LAXPC} light curve variability varies from $7.41\%$ to $15.89\%$ for LMC X-1, while it is in the range 9.7\% - 23.9\% for LMC X-3 (see Table \ref{table2}). We observe that on long time-scales the value of HR remains constant for LMC X-1, while it has a significant variation with a periodic pattern for LMC X-3. From Figure \ref{fig2} we find that LMC X-1 does not show any periodic nature in its light curve, while LMC X-3 does. The light curve periodicity of LMC X-3 during the initial $700$ days is $\sim100 - 200$ days, which is similar to that reported earlier by \citealt{1991ApJ...381..526C,2001MNRAS.320..316N,2012ApJ...756..146S}. However, since MJD 58100 the variability pattern becomes random and do not follow the periodic nature (see right panel of Figure \ref{fig2}).  \par
We explored the temporal properties of the source by generating the PDS. For all epochs of LMC X-1 and LMC X-3, the PDS shows only weak power-law nature as evident from the left and right panels of Figure \ref{fig3}. The fractional rms amplitude varies in between $9\%$ and $16.7\%$ for LMC X-1 and $7.16\%-10.9\%$ for LMC X-3 (see Table \ref{table2}). We also did not observe any QPOs during the different epochs, even though previous publications have reported presence of mHz QPOs in LMC X-1 \citep{2014MNRAS.445.4259A} and LMC X-3 \citep{2000ApJ...542L.127B} in LHS. The weak power-law nature of the PDS, fractional rms and absence of QPOs support the fact that the sources have characteristics of a soft state. Thus from both spectral and temporal characteristics, we understand that the sources remain in a thermal emission dominated spectral state.\par
The spectral fits performed for the broadband {\it AstroSat} observations of the sources are very well described by a thermal disc emission, with presence of occasional steep power-law of photon index $\sim 2.4 - 3.2$ (see Figure \ref{fig4} and Table \ref{table3}). From Table \ref{table3}, we find that the disc temperature remains around $1$ keV and the fractional disc flux contribution is observed to change between $79\%$ and $94\%$ during the different epochs for LMC X-1. We also find that the hardness ratio varies from $0.16$ to $0.34$ during the epochs. This suggests that source spectra are dominated by thermal disc component during all the five epochs. For LMC X-3, we observe that the disc temperature is $\sim1$ keV, fractional disc flux is $>96\%$ and HR is $0.10 - 0.19$ (Table \ref{table3}) during the different epochs. These variations in the spectral parameters indicate that the source LMC X-3 occupied a disc dominated high soft state during the {\it AstroSat} observations. The values of disc parameters obtained are consistent with that obtained by the \textit{BeppoSAX} \citep{2000AdSpR..25..437T, 2001ApJS..133..187H} and \textit{RXTE} observations \citep{2001MNRAS.320..316N}. The bolometric luminosity estimated of $0.07 - 0.10$ $L_{Edd}$ and $0.06 - 0.14$ $L_{Edd}$ suggests that the sources LMC X-1 and LMC X-3 are emitting in sub-Eddington. We also did attempt to find the `true' radius of the accretion disc following \citealt{1998PASJ...50..667K} and got a value of $29.65- 39.17$ km for LMC X-1 and $32.40-37.2$ km for LMC X-3, which are within $\pm3$ km of the mean value. We understand that the radius varies over a long time period for both sources. It has to be noted that these estimations of radius do not follow in sync\footnote{Considering $R_{ISCO} = 9$ km for a non-rotating BH of mass $1 M_{\odot}$ and $3$ km for an extremely rotating BH with same mass, where $R_{ISCO}$ is the radius of innermost stable circular orbit \citep{1983bhwd.book.....S,2006ARA&A..44...49R}} with the source mass estimated (see section \ref{Mna}) and reported earlier \citep{2009ApJ...697..573O,2014ApJ...794..154O}. We did not observe any strong Fe K$_{\alpha}$ line in both sources. But a reflection edge was seen which possibly has a minimal contribution to the total unabsorbed flux. 

Further, we attempt to constrain the source parameters: mass and spin. Since there were no Fe K$_{\alpha}$ emission line detected, we have chosen the continuum fitting method over the entire broadband {\it AstroSat} energy range considered. By using the {\it kerrd} model, we were able to get acceptable spectral fit parameters (mass of $12.64 - 13.00$ $M_{\odot}$) during the different epochs of LMC X-1 but not for any of the observations of LMC X-3. This suggests that the source LMC X-1 belongs to the category of extremely rotating BHs, while LMC X-3 is a weakly rotating source. Using the {\it kerrbb} and {\it simpl} model combination, we were able to constrain both sources' mass, spin and accretion rate. In order to get better estimations on errors of these parameters, we have performed MCMC chain simulation using Goodman-Weare algorithm \citep{2010CAMCS...5...65G} within \textit{XSpec}. Errors estimated with  90\% confidence using this chain yielded a better estimation. We cross-check these error values with the error estimation using MCMC hammer algorithm (Figure \ref{fig6}). 
 We find that the mass for LMC X-1 to be in the range of $7.64^{+0.99}_{-0.25}$ $M_{\odot}$ to $10.00^{+0.52}_{-1.22}$ $M_{\odot}$ and for LMC X-3 it is $5.35^{+0.34}_{-0.34}$ $M_{\odot}$ to $6.22^{+0.48}_{-1.74}$ $M_{\odot}$. All these estimates are in close agreement with the dynamical mass estimate of $10.91\pm1.41$ $M_{\odot}$ and $6.98\pm{0.56}$ $M_{\odot}$ reported by \citep{2009ApJ...697..573O, 2014ApJ...794..154O} for LMC X-1 and LMC X-3 respectively. Similarly, the spin estimate of $0.82^{+0.10}_{-0.02} - 0.92^{+0.04}_{-0.05}$ for LMC X-1 agrees with the previous report by \citealt{2009ApJ...701.1076G,2020ApJ...897...84T}.
\cite{2020MNRAS.498.4404M} has recently reported a spin value of 0.93 by analyzing the first two \textit{AstroSat} observations of LMC X-1, but by considering only the Galactic abundance and a higher spectral hardening factor (1.7). For LMC X-3, the spin estimated is $0.22 \pm 0.03 - 0.41 \pm 0.02$ within 90\% confidence interval (see section \ref{Mna}, Figure \ref{fig6} and Table \ref{table4}), which is better constrained w.r.t. that already reported ($0.25^{+0.20}_{-0.29}$, \citealt{2010ApJ...718L.117S}). We also check the rotating nature of the sources following \citealt{2000ApJ...535..632M}, which expresses the source mass as $M_{x} = R_{in}/8.86\alpha$. Here, $\alpha$ is the positive parameter which has a value of 1 for non-rotating BHs and 1/6 for BHs with extreme rotation. Considering the value of the `true' radius of the accretion disc and the average source mass estimated (see section \ref{SPX1}, \ref{SPX3}, \ref{Mna}), we find the value of $\alpha$ to be $0.37 - 0.50$ for LMC X-1 and $0.63 - 0.73$ for LMC X-3. These estimates of $\alpha$ suggest that the source LMC X-1 probably is a maximally rotating `hole', which is also supported by the spin estimation of $0.82 - 0.92$. Also, the value of $\alpha$ being close to 1 for LMC X-3 suggests that the compact object could be a weakly rotating BH. This fact is consistent with the estimate of spin using continuum fitting method as $0.22 - 0.41$.\par
LMC X-1 and LMC X-3 are peculiar when compared to other persistent BH-XRBs in term of the spectral behavior. Based on the study of spectral and temporal evolution of the sources using {\it AstroSat} observations discussed in this paper, we observe the sources remain in a thermally dominated state over a long duration. We observe that LMC X-3 exists in a HSS during all the {\it AstroSat} observations conducted till date. Previous publications have mentioned a state transition to LHS, which probably suggest frequent monitoring observations of the source are required to probe the accretion dynamics. Similar kind of spectral state transitions over long time scales have been observed for other persistent sources like Cyg X-1 \citep{1999MNRAS.309..496G}, GRS 1758-258 and 1E 1740.7-2942 \citep{1999ApJ...525..901M} using {\it RXTE} observations.

Based on the spectral and temporal studies for the sources LMC X-1 and LMC X-3 using {\it AstroSat} observations, we summarize our results with the following conclusions:
\begin{itemize}
	\item The study of {\it MAXI} light curves over a duration of $4.5$ years show that LMC X-1 has a long term fractional variability of $25\%$, while LMC X-3 has a higher variability of $53\%$.
	\item Long term light curve periodicity of $\sim100-200$ days seen for LMC X-3 during the initial $700$ days, while the variability pattern becomes random later without having any periodic nature.
	\item LMC X-1 is moderately variable from $7.4\% - 16\%$ over short time scale, while LMC X-3 has variability ranging from $9.7\% - 24\%$ during the different observation epochs.
	\item The spectral characteristics suggest the sources have a thermal disc dominated energy spectra.
	\item The weak power-law nature of PDS and evolution of fractional rms amplitude with the absence of low-frequency QPOs supports that sources remained in the thermally dominated soft state.
	\item We constrain the source mass in the range of $7.64 - 10.00$ $M_{\odot}$ for LMC X-1, and $5.35 - 6.22$ $M_{\odot}$ for LMC X-3, which are in close agreement with that already reported.
	\item The spin parameter is estimated to be $0.82-0.92$ for LMC X-1 and is consistent with previous publications. In case of LMC X-3, we could obtain a better constrain of the spin as $0.22 - 0.41$ in contrast with that reported earlier.
\end{itemize}

\section*{Acknowledgements}
\hspace{1cm}
The authors thank the anonymous reviewer for valuable suggestions which have helped in improving this manuscript. We acknowledge the financial support of Indian Space Research Organization (ISRO) under \textit{AstroSat} archival data utilization program Sanction order No. DS-2B-13013(2)/13/2019-Sec.2. This publication uses data from the \textit{AstroSat} mission of the ISRO archived at the Indian Space Science Data Centre (ISSDC). This work has been performed utilizing the calibration databases and auxiliary analysis tools developed, maintained and distributed by {\it AstroSat-SXT} team with members from various institutions in India and abroad. This research has made use of {\it MAXI} data provided by RIKEN, JAXA and the {\it MAXI} team. Also this research made use of software provided by the High Energy Astrophysics Science Archive Research Center (HEASARC) and NASA’s Astrophysics Data System Bibliographic Services. VKA, AN also thank GH, SAG, DD, PDMSA and Director, URSC for encouragement and continuous support to carry out this research.
\par
\noindent{Facilities: {\it AstroSat}, {\it MAXI}.}
\section*{Data Availability}
The data used for analysis in this article are available in {\it AstroSat}-ISSDC website (\url{https://astrobrowse.issdc.gov.in/astro_archive/archive/Home.jsp}) and {\it MAXI} website (\url{http://maxi.riken.jp/top/index.html}).

\bibliographystyle{mnras}
\bibliography{list} 

\begin{thebibliography}{}
\makeatletter
\relax
\def\mn@urlcharsother{\let\do\@makeother \do\$\do\&\do\#\do\^\do\_\do\%\do\~}
\def\mn@doi{\begingroup\mn@urlcharsother \@ifnextchar [ {\mn@doi@}
  {\mn@doi@[]}}
\def\mn@doi@[#1]#2{\def\@tempa{#1}\ifx\@tempa\@empty \href
  {http://dx.doi.org/#2} {doi:#2}\else \href {http://dx.doi.org/#2} {#1}\fi
  \endgroup}
\def\mn@eprint#1#2{\mn@eprint@#1:#2::\@nil}
\def\mn@eprint@arXiv#1{\href {http://arxiv.org/abs/#1} {{\tt arXiv:#1}}}
\def\mn@eprint@dblp#1{\href {http://dblp.uni-trier.de/rec/bibtex/#1.xml}
  {dblp:#1}}
\def\mn@eprint@#1:#2:#3:#4\@nil{\def\@tempa {#1}\def\@tempb {#2}\def\@tempc
  {#3}\ifx \@tempc \@empty \let \@tempc \@tempb \let \@tempb \@tempa \fi \ifx
  \@tempb \@empty \def\@tempb {arXiv}\fi \@ifundefined
  {mn@eprint@\@tempb}{\@tempb:\@tempc}{\expandafter \expandafter \csname
  mn@eprint@\@tempb\endcsname \expandafter{\@tempc}}}

\bibitem[\protect\citeauthoryear{{Agrawal}}{{Agrawal}}{2001}]{2001ASPC..251..512A}
{Agrawal} P.~C.,  2001, in {Inoue} H.,  {Kunieda} H.,  eds,  Astronomical
  Society of the Pacific Conference Series Vol. 251, New Century of X-ray
  Astronomy. p.~512

\bibitem[\protect\citeauthoryear{{Agrawal}, {Nandi}, {Girish}  \&
  {Ramadevi}}{{Agrawal} et~al.}{2018}]{2018MNRAS.477.5437A}
{Agrawal} V.~K.,  {Nandi} A.,  {Girish} V.,   {Ramadevi} M.~C.,  2018, \mn@doi
  [\mnras] {10.1093/mnras/sty1005}, \href
  {https://ui.adsabs.harvard.edu/abs/2018MNRAS.477.5437A} {477, 5437}

\bibitem[\protect\citeauthoryear{{Alam}, {Dewangan}, {Belloni}, {Mukherjee}  \&
  {Jhingan}}{{Alam} et~al.}{2014}]{2014MNRAS.445.4259A}
{Alam} M.~S.,  {Dewangan} G.~C.,  {Belloni} T.,  {Mukherjee} D.,   {Jhingan}
  S.,  2014, \mn@doi [\mnras] {10.1093/mnras/stu2048}, \href
  {https://ui.adsabs.harvard.edu/abs/2014MNRAS.445.4259A} {445, 4259}

\bibitem[\protect\citeauthoryear{{Aneesha}, {Mandal}  \& {Sreehari}}{{Aneesha}
  et~al.}{2019}]{2019MNRAS.486.2705A}
{Aneesha} U.,  {Mandal} S.,   {Sreehari} H.,  2019, \mn@doi [\mnras]
  {10.1093/mnras/stz1000}, \href
  {https://ui.adsabs.harvard.edu/abs/2019MNRAS.486.2705A} {486, 2705}

\bibitem[\protect\citeauthoryear{{Antia} et~al.,}{{Antia}
  et~al.}{2017}]{2017caantialibration}
{Antia} H.~M.,  et~al., 2017, \mn@doi [\apjs] {10.3847/1538-4365/aa7a0e}, \href
  {https://ui.adsabs.harvard.edu/abs/2017ApJS..231...10A} {231, 10}

\bibitem[\protect\citeauthoryear{{Arnaud}}{{Arnaud}}{1996}]{1996ASPC..101...17A}
{Arnaud} K.~A.,  1996, in {Jacoby} G.~H.,  {Barnes} J.,  eds,  Astronomical
  Society of the Pacific Conference Series Vol. 101, Astronomical Data Analysis
  Software and Systems V. p.~17

\bibitem[\protect\citeauthoryear{{Baby}, {Agrawal}, {C}, {Katoch}, {Antia},
  {Mandal}  \& {Nand i}}{{Baby} et~al.}{2020}]{2020MNRAS.tmp.2075B}
{Baby} B.~E.,  {Agrawal} V.~K.,  {C} R.~M.,  {Katoch} T.,  {Antia} H.~M.,
  {Mandal} S.,   {Nand i} A.,  2020, \mn@doi [\mnras] {10.1093/mnras/staa1965},
  \href {https://ui.adsabs.harvard.edu/abs/2020MNRAS.tmp.2075B} {}

\bibitem[\protect\citeauthoryear{{Belloni}}{{Belloni}}{2005}]{2005AIPC..797..197B}
{Belloni} T.,  2005, in {Burderi} L.,  {Antonelli} L.~A.,  {D'Antona} F.,  {di
  Salvo} T.,  {Israel} G.~L.,  {Piersanti} L.,  {Tornamb{\`e}} A.,
  {Straniero} O.,  eds,  American Institute of Physics Conference Series Vol.
  797, Interacting Binaries: Accretion, Evolution, and Outcomes. pp 197--204
  (\mn@eprint {arXiv} {astro-ph/0504185}), \mn@doi{10.1063/1.2130233}

\bibitem[\protect\citeauthoryear{{Belloni} \& {Hasinger}}{{Belloni} \&
  {Hasinger}}{1990}]{1990A&A...230..103B}
{Belloni} T.,  {Hasinger} G.,  1990, \aap, \href
  {https://ui.adsabs.harvard.edu/abs/1990A&A...230..103B} {230, 103}

\bibitem[\protect\citeauthoryear{{Belloni}, {Motta}  \&
  {Mu{\~n}oz-Darias}}{{Belloni} et~al.}{2011}]{2011BASI...39..409B}
{Belloni} T.~M.,  {Motta} S.~E.,   {Mu{\~n}oz-Darias} T.,  2011, Bulletin of
  the Astronomical Society of India, \href
  {https://ui.adsabs.harvard.edu/abs/2011BASI...39..409B} {39, 409}

\bibitem[\protect\citeauthoryear{{Bevington} \& {Robinson}}{{Bevington} \&
  {Robinson}}{2003}]{2003drea.book.....B}
{Bevington} P.~R.,  {Robinson} D.~K.,  2003, {Data reduction and error analysis
  for the physical sciences}

\bibitem[\protect\citeauthoryear{{Boller} et~al.,}{{Boller}
  et~al.}{2002}]{2002MNRAS.329L...1B}
{Boller} T.,  et~al., 2002, \mn@doi [\mnras]
  {10.1046/j.1365-8711.2002.05040.x}, \href
  {https://ui.adsabs.harvard.edu/abs/2002MNRAS.329L...1B} {329, L1}

\bibitem[\protect\citeauthoryear{{Boyd}, {Smale}, {Homan}, {Jonker}, {van der
  Klis}  \& {Kuulkers}}{{Boyd} et~al.}{2000}]{2000ApJ...542L.127B}
{Boyd} P.~T.,  {Smale} A.~P.,  {Homan} J.,  {Jonker} P.~G.,  {van der Klis} M.,
    {Kuulkers} E.,  2000, \mn@doi [\apjl] {10.1086/312931}, \href
  {https://ui.adsabs.harvard.edu/abs/2000ApJ...542L.127B} {542, L127}

\bibitem[\protect\citeauthoryear{{Casella}, {Belloni}  \& {Stella}}{{Casella}
  et~al.}{2005}]{2005ApJ...629..403C}
{Casella} P.,  {Belloni} T.,   {Stella} L.,  2005, \mn@doi [\apj]
  {10.1086/431174}, \href
  {https://ui.adsabs.harvard.edu/abs/2005ApJ...629..403C} {629, 403}

\bibitem[\protect\citeauthoryear{{Chakrabarti} \& {Titarchuk}}{{Chakrabarti} \&
  {Titarchuk}}{1995}]{1995ApJ...455..623C}
{Chakrabarti} S.,  {Titarchuk} L.~G.,  1995, \mn@doi [\apj] {10.1086/176610},
  \href {https://ui.adsabs.harvard.edu/abs/1995ApJ...455..623C} {455, 623}

\bibitem[\protect\citeauthoryear{{Chen} \& {Taam}}{{Chen} \&
  {Taam}}{1996}]{1996ApJ...466..404C}
{Chen} X.,  {Taam} R.~E.,  1996, \mn@doi [\apj] {10.1086/177519}, \href
  {https://ui.adsabs.harvard.edu/abs/1996ApJ...466..404C} {466, 404}

\bibitem[\protect\citeauthoryear{{Chen}, {Shrader}  \& {Livio}}{{Chen}
  et~al.}{1997}]{1997ApJ...491..312C}
{Chen} W.,  {Shrader} C.~R.,   {Livio} M.,  1997, \mn@doi [\apj]
  {10.1086/304921}, \href
  {https://ui.adsabs.harvard.edu/abs/1997ApJ...491..312C} {491, 312}

\bibitem[\protect\citeauthoryear{{Corral-Santana}, {Casares},
  {Mu{\~n}oz-Darias}, {Bauer}, {Mart{\'\i}nez-Pais}  \&
  {Russell}}{{Corral-Santana} et~al.}{2016}]{2016A&A...587A..61C}
{Corral-Santana} J.~M.,  {Casares} J.,  {Mu{\~n}oz-Darias} T.,  {Bauer} F.~E.,
  {Mart{\'\i}nez-Pais} I.~G.,   {Russell} D.~M.,  2016, \mn@doi [\aap]
  {10.1051/0004-6361/201527130}, \href
  {https://ui.adsabs.harvard.edu/abs/2016A&A...587A..61C} {587, A61}

\bibitem[\protect\citeauthoryear{{Cowley} et~al.,}{{Cowley}
  et~al.}{1991}]{1991ApJ...381..526C}
{Cowley} A.~P.,  et~al., 1991, \mn@doi [\apj] {10.1086/170676}, \href
  {https://ui.adsabs.harvard.edu/abs/1991ApJ...381..526C} {381, 526}

\bibitem[\protect\citeauthoryear{{Cowley}, {Schmidtke}, {Anderson}  \&
  {McGrath}}{{Cowley} et~al.}{1995}]{1995PASP..107..145C}
{Cowley} A.~P.,  {Schmidtke} P.~C.,  {Anderson} A.~L.,   {McGrath} T.~K.,
  1995, \mn@doi [\pasp] {10.1086/133530}, \href
  {https://ui.adsabs.harvard.edu/abs/1995PASP..107..145C} {107, 145}

\bibitem[\protect\citeauthoryear{{Ebisawa}, {Mitsuda}  \& {Inoue}}{{Ebisawa}
  et~al.}{1989}]{1989PASJ...41..519E}
{Ebisawa} K.,  {Mitsuda} K.,   {Inoue} H.,  1989, \pasj, \href
  {https://ui.adsabs.harvard.edu/abs/1989PASJ...41..519E} {41, 519}

\bibitem[\protect\citeauthoryear{{Ebisawa}, {Makino}, {Mitsuda}, {Belloni},
  {Cowley}, {Schmidtke}  \& {Treves}}{{Ebisawa}
  et~al.}{1993}]{1993ApJ...403..684E}
{Ebisawa} K.,  {Makino} F.,  {Mitsuda} K.,  {Belloni} T.,  {Cowley} A.~P.,
  {Schmidtke} P.~C.,   {Treves} A.,  1993, \mn@doi [\apj] {10.1086/172239},
  \href {https://ui.adsabs.harvard.edu/abs/1993ApJ...403..684E} {403, 684}

\bibitem[\protect\citeauthoryear{{Fabian} \& {Vaughan}}{{Fabian} \&
  {Vaughan}}{2003}]{2003MNRAS.340L..28F}
{Fabian} A.~C.,  {Vaughan} S.,  2003, \mn@doi [\mnras]
  {10.1046/j.1365-8711.2003.06465.x}, \href
  {https://ui.adsabs.harvard.edu/abs/2003MNRAS.340L..28F} {340, L28}

\bibitem[\protect\citeauthoryear{{Fabian} et~al.,}{{Fabian}
  et~al.}{2002}]{2002MNRAS.335L...1F}
{Fabian} A.~C.,  et~al., 2002, \mn@doi [\mnras]
  {10.1046/j.1365-8711.2002.05740.x}, \href
  {https://ui.adsabs.harvard.edu/abs/2002MNRAS.335L...1F} {335, L1}

\bibitem[\protect\citeauthoryear{{Foreman-Mackey}, {Hogg}, {Lang}  \&
  {Goodman}}{{Foreman-Mackey} et~al.}{2013}]{2013PASP..125..306F}
{Foreman-Mackey} D.,  {Hogg} D.~W.,  {Lang} D.,   {Goodman} J.,  2013, \mn@doi
  [\pasp] {10.1086/670067}, \href
  {https://ui.adsabs.harvard.edu/abs/2013PASP..125..306F} {125, 306}

\bibitem[\protect\citeauthoryear{{Gierli{\'n}ski}, {Zdziarski}, {Poutanen},
  {Coppi}, {Ebisawa}  \& {Johnson}}{{Gierli{\'n}ski}
  et~al.}{1999}]{1999MNRAS.309..496G}
{Gierli{\'n}ski} M.,  {Zdziarski} A.~A.,  {Poutanen} J.,  {Coppi} P.~S.,
  {Ebisawa} K.,   {Johnson} W.~N.,  1999, \mn@doi [\mnras]
  {10.1046/j.1365-8711.1999.02875.x}, \href
  {https://ui.adsabs.harvard.edu/abs/1999MNRAS.309..496G} {309, 496}

\bibitem[\protect\citeauthoryear{{Goodman} \& {Weare}}{{Goodman} \&
  {Weare}}{2010}]{2010CAMCS...5...65G}
{Goodman} J.,  {Weare} J.,  2010, \mn@doi [Communications in Applied
  Mathematics and Computational Science] {10.2140/camcos.2010.5.65}, \href
  {https://ui.adsabs.harvard.edu/abs/2010CAMCS...5...65G} {5, 65}

\bibitem[\protect\citeauthoryear{{Gou} et~al.,}{{Gou}
  et~al.}{2009}]{2009ApJ...701.1076G}
{Gou} L.,  et~al., 2009, \mn@doi [\apj] {10.1088/0004-637X/701/2/1076}, \href
  {https://ui.adsabs.harvard.edu/abs/2009ApJ...701.1076G} {701, 1076}

\bibitem[\protect\citeauthoryear{{Haardt} et~al.,}{{Haardt}
  et~al.}{2001}]{2001ApJS..133..187H}
{Haardt} F.,  et~al., 2001, \mn@doi [\apjs] {10.1086/319186}, \href
  {https://ui.adsabs.harvard.edu/abs/2001ApJS..133..187H} {133, 187}

\bibitem[\protect\citeauthoryear{{Hanke}, {Wilms}, {Nowak}, {Barrag{\'a}n}  \&
  {Schulz}}{{Hanke} et~al.}{2010}]{hanke10}
{Hanke} M.,  {Wilms} J.,  {Nowak} M.~A.,  {Barrag{\'a}n} L.,   {Schulz} N.~S.,
  2010, \mn@doi [\aap] {10.1051/0004-6361/200913583}, \href
  {https://ui.adsabs.harvard.edu/abs/2010A%26A...509L...8H} {509, L8}

\bibitem[\protect\citeauthoryear{{Homan}, {Wijnands}, {van der Klis},
  {Belloni}, {van Paradijs}, {Klein-Wolt}, {Fender}  \& {M{\'e}ndez}}{{Homan}
  et~al.}{2001}]{2001ApJS..132..377H}
{Homan} J.,  {Wijnands} R.,  {van der Klis} M.,  {Belloni} T.,  {van Paradijs}
  J.,  {Klein-Wolt} M.,  {Fender} R.,   {M{\'e}ndez} M.,  2001, \mn@doi [\apjs]
  {10.1086/318954}, \href
  {https://ui.adsabs.harvard.edu/abs/2001ApJS..132..377H} {132, 377}

\bibitem[\protect\citeauthoryear{{Katoch}, {Blessy E.}, {Nandi}, {Agrawal},
  {Antia}  \& {Mukerjee}}{{Katoch} et~al.}{2020}]{Katoch2020}
{Katoch} T.,  {Blessy E.} B.,  {Nandi} A.,  {Agrawal} V.~K.,  {Antia} H.~M.,
  {Mukerjee} M.,  2020

\bibitem[\protect\citeauthoryear{{Kubota}, {Tanaka}, {Makishima}, {Ueda},
  {Dotani}, {Inoue}  \& {Yamaoka}}{{Kubota} et~al.}{1998}]{1998PASJ...50..667K}
{Kubota} A.,  {Tanaka} Y.,  {Makishima} K.,  {Ueda} Y.,  {Dotani} T.,  {Inoue}
  H.,   {Yamaoka} K.,  1998, \mn@doi [\pasj] {10.1093/pasj/50.6.667}, \href
  {https://ui.adsabs.harvard.edu/abs/1998PASJ...50..667K} {50, 667}

\bibitem[\protect\citeauthoryear{{Li}, {Zimmerman}, {Narayan}  \&
  {McClintock}}{{Li} et~al.}{2005}]{2005ApJS..157..335L}
{Li} L.-X.,  {Zimmerman} E.~R.,  {Narayan} R.,   {McClintock} J.~E.,  2005,
  \mn@doi [\apjs] {10.1086/428089}, \href
  {https://ui.adsabs.harvard.edu/abs/2005ApJS..157..335L} {157, 335}

\bibitem[\protect\citeauthoryear{{Main}, {Smith}, {Heindl}, {Swank},
  {Leventhal}, {Mirabel}  \& {Rodr{\'\i}guez}}{{Main}
  et~al.}{1999}]{1999ApJ...525..901M}
{Main} D.~S.,  {Smith} D.~M.,  {Heindl} W.~A.,  {Swank} J.,  {Leventhal} M.,
  {Mirabel} I.~F.,   {Rodr{\'\i}guez} L.~F.,  1999, \mn@doi [\apj]
  {10.1086/307935}, \href
  {https://ui.adsabs.harvard.edu/abs/1999ApJ...525..901M} {525, 901}

\bibitem[\protect\citeauthoryear{{Makishima} et~al.,}{{Makishima}
  et~al.}{2000}]{2000ApJ...535..632M}
{Makishima} K.,  et~al., 2000, \mn@doi [\apj] {10.1086/308868}, \href
  {https://ui.adsabs.harvard.edu/abs/2000ApJ...535..632M} {535, 632}

\bibitem[\protect\citeauthoryear{Mark, Price, Rodrigues, Seward  \& Swift}{Mark
  et~al.}{1969}]{mark1969}
Mark H.,  Price R.,  Rodrigues R.,  Seward F.~D.,   Swift C.~D.,  1969, \mn@doi
  [\apjl] {10.1086/180322}, \href
  {https://ui.adsabs.harvard.edu/abs/1969ApJ...155L.143M} {155, L143}

\bibitem[\protect\citeauthoryear{{McClintock} \& {Remillard}}{{McClintock} \&
  {Remillard}}{1986}]{1986ApJ...308..110M}
{McClintock} J.~E.,  {Remillard} R.~A.,  1986, \mn@doi [\apj] {10.1086/164482},
  \href {https://ui.adsabs.harvard.edu/abs/1986ApJ...308..110M} {308, 110}

\bibitem[\protect\citeauthoryear{{McClintock}, {Shafee}, {Narayan},
  {Remillard}, {Davis}  \& {Li}}{{McClintock}
  et~al.}{2006}]{2006ApJ...652..518M}
{McClintock} J.~E.,  {Shafee} R.,  {Narayan} R.,  {Remillard} R.~A.,  {Davis}
  S.~W.,   {Li} L.-X.,  2006, \mn@doi [\apj] {10.1086/508457}, \href
  {https://ui.adsabs.harvard.edu/abs/2006ApJ...652..518M} {652, 518}

\bibitem[\protect\citeauthoryear{{McClintock} et~al.,}{{McClintock}
  et~al.}{2011}]{2011CQGra..28k4009M}
{McClintock} J.~E.,  et~al., 2011, \mn@doi [Classical and Quantum Gravity]
  {10.1088/0264-9381/28/11/114009}, \href
  {https://ui.adsabs.harvard.edu/abs/2011CQGra..28k4009M} {28, 114009}

\bibitem[\protect\citeauthoryear{{McClintock}, {Narayan}  \&
  {Steiner}}{{McClintock} et~al.}{2014}]{2014SSRv..183..295M}
{McClintock} J.~E.,  {Narayan} R.,   {Steiner} J.~F.,  2014, \mn@doi [\ssr]
  {10.1007/s11214-013-0003-9}, \href
  {https://ui.adsabs.harvard.edu/abs/2014SSRv..183..295M} {183, 295}

\bibitem[\protect\citeauthoryear{{Mudambi}, {Rao}, {Gudennavar}, {Misra}  \&
  {Bubbly}}{{Mudambi} et~al.}{2020}]{2020MNRAS.498.4404M}
{Mudambi} S.~P.,  {Rao} A.,  {Gudennavar} S.~B.,  {Misra} R.,   {Bubbly} S.~G.,
   2020, \mn@doi [\mnras] {10.1093/mnras/staa2656}, \href
  {https://ui.adsabs.harvard.edu/abs/2020MNRAS.498.4404M} {498, 4404}

\bibitem[\protect\citeauthoryear{{Nandi}, {Debnath}, {Mandal}  \&
  {Chakrabarti}}{{Nandi} et~al.}{2012}]{2012A&A...542A..56N}
{Nandi} A.,  {Debnath} D.,  {Mandal} S.,   {Chakrabarti} S.~K.,  2012, \mn@doi
  [\aap] {10.1051/0004-6361/201117844}, \href
  {https://ui.adsabs.harvard.edu/abs/2012A&A...542A..56N} {542, A56}

\bibitem[\protect\citeauthoryear{{Nandi} et~al.,}{{Nandi}
  et~al.}{2018}]{2018Ap&SS.363...90N}
{Nandi} A.,  et~al., 2018, \mn@doi [\apss] {10.1007/s10509-018-3314-1}, \href
  {https://ui.adsabs.harvard.edu/abs/2018Ap&SS.363...90N} {363, 90}

\bibitem[\protect\citeauthoryear{{Nowak}}{{Nowak}}{1995}]{1995PASP..107.1207N}
{Nowak} M.~A.,  1995, \mn@doi [\pasp] {10.1086/133679}, \href
  {https://ui.adsabs.harvard.edu/abs/1995PASP..107.1207N} {107, 1207}

\bibitem[\protect\citeauthoryear{{Nowak}, {Wilms}, {Heindl}, {Pottschmidt},
  {Dove}  \& {Begelman}}{{Nowak} et~al.}{2001}]{2001MNRAS.320..316N}
{Nowak} M.~A.,  {Wilms} J.,  {Heindl} W.~A.,  {Pottschmidt} K.,  {Dove} J.~B.,
   {Begelman} M.~C.,  2001, \mn@doi [\mnras]
  {10.1046/j.1365-8711.2001.03984.x}, \href
  {https://ui.adsabs.harvard.edu/abs/2001MNRAS.320..316N} {320, 316}

\bibitem[\protect\citeauthoryear{{Orosz} et~al.,}{{Orosz}
  et~al.}{2009}]{2009ApJ...697..573O}
{Orosz} J.~A.,  et~al., 2009, \mn@doi [\apj] {10.1088/0004-637X/697/1/573},
  \href {https://ui.adsabs.harvard.edu/abs/2009ApJ...697..573O} {697, 573}

\bibitem[\protect\citeauthoryear{{Orosz}, {Steiner}, {McClintock}, {Buxton},
  {Bailyn}, {Steeghs}, {Guberman}  \& {Torres}}{{Orosz}
  et~al.}{2014}]{2014ApJ...794..154O}
{Orosz} J.~A.,  {Steiner} J.~F.,  {McClintock} J.~E.,  {Buxton} M.~M.,
  {Bailyn} C.~D.,  {Steeghs} D.,  {Guberman} A.,   {Torres} M. A.~P.,  2014,
  \mn@doi [\apj] {10.1088/0004-637X/794/2/154}, \href
  {https://ui.adsabs.harvard.edu/abs/2014ApJ...794..154O} {794, 154}

\bibitem[\protect\citeauthoryear{{Radhika} \& {Nandi}}{{Radhika} \&
  {Nandi}}{2014}]{2014AdSpR..54.1678R}
{Radhika} D.,  {Nandi} A.,  2014, \mn@doi [Advances in Space Research]
  {10.1016/j.asr.2014.06.039}, \href
  {https://ui.adsabs.harvard.edu/abs/2014AdSpR..54.1678R} {54, 1678}

\bibitem[\protect\citeauthoryear{{Radhika}, {Nandi}, {Agrawal}  \&
  {Seetha}}{{Radhika} et~al.}{2016a}]{2016MNRAS.460.4403R}
{Radhika} D.,  {Nandi} A.,  {Agrawal} V.~K.,   {Seetha} S.,  2016a, \mn@doi
  [\mnras] {10.1093/mnras/stw1239}, \href
  {https://ui.adsabs.harvard.edu/abs/2016MNRAS.460.4403R} {460, 4403}

\bibitem[\protect\citeauthoryear{{Radhika}, {Nandi}, {Agrawal}  \&
  {Mandal}}{{Radhika} et~al.}{2016b}]{2016MNRAS.462.1834R}
{Radhika} D.,  {Nandi} A.,  {Agrawal} V.~K.,   {Mandal} S.,  2016b, \mn@doi
  [\mnras] {10.1093/mnras/stw1755}, \href
  {https://ui.adsabs.harvard.edu/abs/2016MNRAS.462.1834R} {462, 1834}

\bibitem[\protect\citeauthoryear{{Radhika}, {Sreehari}, {Nandi}, {Iyer}  \&
  {Mandal}}{{Radhika} et~al.}{2018}]{2018Ap&SS.363..189D}
{Radhika} D.,  {Sreehari} H.,  {Nandi} A.,  {Iyer} N.,   {Mandal} S.,  2018,
  \mn@doi [\apss] {10.1007/s10509-018-3411-1}, \href
  {https://ui.adsabs.harvard.edu/abs/2018Ap&SS.363..189D} {363, 189}

\bibitem[\protect\citeauthoryear{{Remillard} \& {McClintock}}{{Remillard} \&
  {McClintock}}{2006}]{2006ARA&A..44...49R}
{Remillard} R.~A.,  {McClintock} J.~E.,  2006, \mn@doi [\araa]
  {10.1146/annurev.astro.44.051905.092532}, \href
  {https://ui.adsabs.harvard.edu/abs/2006ARA&A..44...49R} {44, 49}

\bibitem[\protect\citeauthoryear{{Schmidtke}, {Ponder}  \&
  {Cowley}}{{Schmidtke} et~al.}{1999}]{1999AJ....117.1292S}
{Schmidtke} P.~C.,  {Ponder} A.~L.,   {Cowley} A.~P.,  1999, \mn@doi [\aj]
  {10.1086/300793}, \href
  {https://ui.adsabs.harvard.edu/abs/1999AJ....117.1292S} {117, 1292}

\bibitem[\protect\citeauthoryear{{Shafee}, {McClintock}, {Narayan}, {Davis},
  {Li}  \& {Remillard}}{{Shafee} et~al.}{2006}]{2006ApJ...636L.113S}
{Shafee} R.,  {McClintock} J.~E.,  {Narayan} R.,  {Davis} S.~W.,  {Li} L.-X.,
  {Remillard} R.~A.,  2006, \mn@doi [\apjl] {10.1086/498938}, \href
  {https://ui.adsabs.harvard.edu/abs/2006ApJ...636L.113S} {636, L113}

\bibitem[\protect\citeauthoryear{{Shakura} \& {Sunyaev}}{{Shakura} \&
  {Sunyaev}}{1973}]{1973A&A....24..337S}
{Shakura} N.~I.,  {Sunyaev} R.~A.,  1973, \aap, \href
  {https://ui.adsabs.harvard.edu/abs/1973A&A....24..337S} {500, 33}

\bibitem[\protect\citeauthoryear{{Shapiro}}{{Shapiro}}{1973}]{1973ApJ...180..531S}
{Shapiro} S.~L.,  1973, \mn@doi [\apj] {10.1086/151982}, \href
  {https://ui.adsabs.harvard.edu/abs/1973ApJ...180..531S} {180, 531}

\bibitem[\protect\citeauthoryear{{Shapiro} \& {Teukolsky}}{{Shapiro} \&
  {Teukolsky}}{1983}]{1983bhwd.book.....S}
{Shapiro} S.~L.,  {Teukolsky} S.~A.,  1983, {Black holes, white dwarfs, and
  neutron stars : the physics of compact objects}

\bibitem[\protect\citeauthoryear{{Singh} et~al.,}{{Singh}
  et~al.}{2017}]{2017JApA...38...29S}
{Singh} K.~P.,  et~al., 2017, \mn@doi [Journal of Astrophysics and Astronomy]
  {10.1007/s12036-017-9448-7}, \href
  {https://ui.adsabs.harvard.edu/abs/2017JApA...38...29S} {38, 29}

\bibitem[\protect\citeauthoryear{{Smale} \& {Boyd}}{{Smale} \&
  {Boyd}}{2012}]{2012ApJ...756..146S}
{Smale} A.~P.,  {Boyd} P.~T.,  2012, \mn@doi [\apj]
  {10.1088/0004-637X/756/2/146}, \href
  {https://ui.adsabs.harvard.edu/abs/2012ApJ...756..146S} {756, 146}

\bibitem[\protect\citeauthoryear{{Soria}, {Wu}, {Page}  \& {Sakelliou}}{{Soria}
  et~al.}{2001}]{2001A&A...365L.273S}
{Soria} R.,  {Wu} K.,  {Page} M.~J.,   {Sakelliou} I.,  2001, \mn@doi [\aap]
  {10.1051/0004-6361:20000065}, \href
  {https://ui.adsabs.harvard.edu/abs/2001A&A...365L.273S} {365, L273}

\bibitem[\protect\citeauthoryear{{Sreehari}, {Nandi}, {Radhika}, {Iyer}  \&
  {Mandal}}{{Sreehari} et~al.}{2018}]{2018JApA...39....5S}
{Sreehari} H.,  {Nandi} A.,  {Radhika} D.,  {Iyer} N.,   {Mandal} S.,  2018,
  \mn@doi [Journal of Astrophysics and Astronomy] {10.1007/s12036-018-9510-0},
  \href {https://ui.adsabs.harvard.edu/abs/2018JApA...39....5S} {39, 5}

\bibitem[\protect\citeauthoryear{{Sreehari}, {Ravishankar}, {Iyer}, {Agrawal},
  {Katoch}, {Mandal}  \& {Nand i}}{{Sreehari}
  et~al.}{2019}]{2019MNRAS.487..928S}
{Sreehari} H.,  {Ravishankar} B.~T.,  {Iyer} N.,  {Agrawal} V.~K.,  {Katoch}
  T.~B.,  {Mandal} S.,   {Nand i} A.,  2019, \mn@doi [\mnras]
  {10.1093/mnras/stz1327}, \href
  {https://ui.adsabs.harvard.edu/abs/2019MNRAS.487..928S} {487, 928}

\bibitem[\protect\citeauthoryear{{Sreehari}, {Nandi}, {Das}, {Agrawal},
  {Mandal}, {Ramadevi}  \& {Katoch}}{{Sreehari}
  et~al.}{2020}]{2020MNRAS.499.5891S}
{Sreehari} H.,  {Nandi} A.,  {Das} S.,  {Agrawal} V.~K.,  {Mandal} S.,
  {Ramadevi} M.~C.,   {Katoch} T.,  2020, \mn@doi [\mnras]
  {10.1093/mnras/staa3135}, \href
  {https://ui.adsabs.harvard.edu/abs/2020MNRAS.499.5891S} {499, 5891}

\bibitem[\protect\citeauthoryear{{Steiner}, {Narayan}, {McClintock}  \&
  {Ebisawa}}{{Steiner} et~al.}{2009}]{2009PASP..121.1279S}
{Steiner} J.~F.,  {Narayan} R.,  {McClintock} J.~E.,   {Ebisawa} K.,  2009,
  \mn@doi [\pasp] {10.1086/648535}, \href
  {https://ui.adsabs.harvard.edu/abs/2009PASP..121.1279S} {121, 1279}

\bibitem[\protect\citeauthoryear{{Steiner}, {McClintock}, {Remillard}, {Gou},
  {Yamada}  \& {Narayan}}{{Steiner} et~al.}{2010}]{2010ApJ...718L.117S}
{Steiner} J.~F.,  {McClintock} J.~E.,  {Remillard} R.~A.,  {Gou} L.,  {Yamada}
  S.,   {Narayan} R.,  2010, \mn@doi [\apjl] {10.1088/2041-8205/718/2/L117},
  \href {https://ui.adsabs.harvard.edu/abs/2010ApJ...718L.117S} {718, L117}

\bibitem[\protect\citeauthoryear{{Steiner}, {McClintock}, {Orosz}, {Remillard},
  {Bailyn}, {Kolehmainen}  \& {Straub}}{{Steiner}
  et~al.}{2014}]{2014ApJ...793L..29S}
{Steiner} J.~F.,  {McClintock} J.~E.,  {Orosz} J.~A.,  {Remillard} R.~A.,
  {Bailyn} C.~D.,  {Kolehmainen} M.,   {Straub} O.,  2014, \mn@doi [\apjl]
  {10.1088/2041-8205/793/2/L29}, \href
  {https://ui.adsabs.harvard.edu/abs/2014ApJ...793L..29S} {793, L29}

\bibitem[\protect\citeauthoryear{{Tanaka} \& {Lewin}}{{Tanaka} \&
  {Lewin}}{1995}]{1995xrbi.nasa..126T}
{Tanaka} Y.,  {Lewin} W.~H.~G.,  1995, X-ray Binaries, \href
  {https://ui.adsabs.harvard.edu/abs/1995xrbi.nasa..126T} {pp 126--174}

\bibitem[\protect\citeauthoryear{{Tanaka} \& {Shibazaki}}{{Tanaka} \&
  {Shibazaki}}{1996}]{1996ARA&A..34..607T}
{Tanaka} Y.,  {Shibazaki} N.,  1996, \mn@doi [\araa]
  {10.1146/annurev.astro.34.1.607}, \href
  {https://ui.adsabs.harvard.edu/abs/1996ARA&A..34..607T} {34, 607}

\bibitem[\protect\citeauthoryear{{Tetarenko}, {Sivakoff}, {Heinke}  \&
  {Gladstone}}{{Tetarenko} et~al.}{2016}]{2016ApJS..222...15T}
{Tetarenko} B.~E.,  {Sivakoff} G.~R.,  {Heinke} C.~O.,   {Gladstone} J.~C.,
  2016, \mn@doi [\apjs] {10.3847/0067-0049/222/2/15}, \href
  {https://ui.adsabs.harvard.edu/abs/2016ApJS..222...15T} {222, 15}

\bibitem[\protect\citeauthoryear{{Torpin}, {Boyd}, {Smale}  \&
  {Valencic}}{{Torpin} et~al.}{2017}]{2017ApJ...849...32T}
{Torpin} T.~J.,  {Boyd} P.~T.,  {Smale} A.~P.,   {Valencic} L.~A.,  2017,
  \mn@doi [\apj] {10.3847/1538-4357/aa8f96}, \href
  {https://ui.adsabs.harvard.edu/abs/2017ApJ...849...32T} {849, 32}

\bibitem[\protect\citeauthoryear{{Treves}, {Belloni}, {Chiappetti}, {Maraschi},
  {Stella}, {Tanzi}  \& {van der Klis}}{{Treves}
  et~al.}{1988}]{1988ApJ...325..119T}
{Treves} A.,  {Belloni} T.,  {Chiappetti} L.,  {Maraschi} L.,  {Stella} L.,
  {Tanzi} E.~G.,   {van der Klis} M.,  1988, \mn@doi [\apj] {10.1086/165987},
  \href {https://ui.adsabs.harvard.edu/abs/1988ApJ...325..119T} {325, 119}

\bibitem[\protect\citeauthoryear{{Treves} et~al.,}{{Treves}
  et~al.}{2000}]{2000AdSpR..25..437T}
{Treves} A.,  et~al., 2000, \mn@doi [Advances in Space Research]
  {10.1016/S0273-1177(99)00775-9}, \href
  {https://ui.adsabs.harvard.edu/abs/2000AdSpR..25..437T} {25, 437}

\bibitem[\protect\citeauthoryear{{Tripathi} et~al.,}{{Tripathi}
  et~al.}{2020}]{2020ApJ...897...84T}
{Tripathi} A.,  et~al., 2020, \mn@doi [\apj] {10.3847/1538-4357/ab9600}, \href
  {https://ui.adsabs.harvard.edu/abs/2020ApJ...897...84T} {897, 84}

\bibitem[\protect\citeauthoryear{{Tsunemi}, {Kitamoto}, {Okamura}  \&
  {Roussel-Dupre}}{{Tsunemi} et~al.}{1989}]{1989ApJ...337L..81T}
{Tsunemi} H.,  {Kitamoto} S.,  {Okamura} S.,   {Roussel-Dupre} D.,  1989,
  \mn@doi [\apjl] {10.1086/185383}, \href
  {https://ui.adsabs.harvard.edu/abs/1989ApJ...337L..81T} {337, L81}

\bibitem[\protect\citeauthoryear{{Vaughan}, {Reeves}, {Warwick}  \&
  {Edelson}}{{Vaughan} et~al.}{1999}]{1999MNRAS.309..113V}
{Vaughan} S.,  {Reeves} J.,  {Warwick} R.,   {Edelson} R.,  1999, \mn@doi
  [\mnras] {10.1046/j.1365-8711.1999.02811.x}, \href
  {https://ui.adsabs.harvard.edu/abs/1999MNRAS.309..113V} {309, 113}

\bibitem[\protect\citeauthoryear{{Vaughan}, {Edelson}, {Warwick}  \&
  {Uttley}}{{Vaughan} et~al.}{2003}]{2003MNRAS.345.1271V}
{Vaughan} S.,  {Edelson} R.,  {Warwick} R.~S.,   {Uttley} P.,  2003, \mn@doi
  [\mnras] {10.1046/j.1365-2966.2003.07042.x}, \href
  {https://ui.adsabs.harvard.edu/abs/2003MNRAS.345.1271V} {345, 1271}

\bibitem[\protect\citeauthoryear{{Wilms}, {Allen}  \& {McCray}}{{Wilms}
  et~al.}{2000}]{2000ApJ...542..914W}
{Wilms} J.,  {Allen} A.,   {McCray} R.,  2000, \mn@doi [\apj] {10.1086/317016},
  \href {https://ui.adsabs.harvard.edu/abs/2000ApJ...542..914W} {542, 914}

\bibitem[\protect\citeauthoryear{{Wilms}, {Nowak}, {Pottschmidt}, {Heindl},
  {Dove}  \& {Begelman}}{{Wilms} et~al.}{2001}]{2001MNRAS.320..327W}
{Wilms} J.,  {Nowak} M.~A.,  {Pottschmidt} K.,  {Heindl} W.~A.,  {Dove} J.~B.,
   {Begelman} M.~C.,  2001, \mn@doi [\mnras]
  {10.1046/j.1365-8711.2001.03983.x}, \href
  {https://ui.adsabs.harvard.edu/abs/2001MNRAS.320..327W} {320, 327}

\bibitem[\protect\citeauthoryear{{Yadav} et~al.,}{{Yadav}
  et~al.}{2016}]{2016SPIE.9905E..1DY}
{Yadav} J.~S.,  et~al., 2016, {Large Area X-ray Proportional Counter (LAXPC)
  instrument onboard ASTROSAT}.
p. 99051D, \mn@doi{10.1117/12.2231857}

\bibitem[\protect\citeauthoryear{{Zdziarski}, {Poutanen}, {Paciesas}  \&
  {Wen}}{{Zdziarski} et~al.}{2002}]{2002ApJ...578..357Z}
{Zdziarski} A.~A.,  {Poutanen} J.,  {Paciesas} W.~S.,   {Wen} L.,  2002,
  \mn@doi [\apj] {10.1086/342402}, \href
  {https://ui.adsabs.harvard.edu/abs/2002ApJ...578..357Z} {578, 357}

\bibitem[\protect\citeauthoryear{{Zhang}, {Cui}, {Harmon}, {Paciesas},
  {Remillard}  \& {van Paradijs}}{{Zhang} et~al.}{1997a}]{1997ApJ...477L..95Z}
{Zhang} S.~N.,  {Cui} W.,  {Harmon} B.~A.,  {Paciesas} W.~S.,  {Remillard}
  R.~E.,   {van Paradijs} J.,  1997a, \mn@doi [\apjl] {10.1086/310530}, \href
  {https://ui.adsabs.harvard.edu/abs/1997ApJ...477L..95Z} {477, L95}

\bibitem[\protect\citeauthoryear{{Zhang}, {Cui}  \& {Chen}}{{Zhang}
  et~al.}{1997b}]{1997ApJ...482L.155Z}
{Zhang} S.~N.,  {Cui} W.,   {Chen} W.,  1997b, \mn@doi [\apjl]
  {10.1086/310705}, \href
  {https://ui.adsabs.harvard.edu/abs/1997ApJ...482L.155Z} {482, L155}

\bibitem[\protect\citeauthoryear{{{\.Z}ycki}, {Done}  \& {Smith}}{{{\.Z}ycki}
  et~al.}{1999}]{1999MNRAS.309..561Z}
{{\.Z}ycki} P.~T.,  {Done} C.,   {Smith} D.~A.,  1999, \mn@doi [\mnras]
  {10.1046/j.1365-8711.1999.02885.x}, \href
  {https://ui.adsabs.harvard.edu/abs/1999MNRAS.309..561Z} {309, 561}

\makeatother
\end{thebibliography}




\appendix

\label{app}
\bsp	
\label{lastpage}
\end{document}